\documentclass[twocolumn,trackchanges]{aastex701}
\usepackage{amsmath}
\usepackage{booktabs}

\newcommand{\hb}{H$\beta$}
\newcommand{\ha}{H$\alpha$}
\newcommand{\oiii}{[O\,{\sc iii}]}
\newcommand{\nii}{[N\,{\sc ii}]}
\newcommand{\sii}{[S\,{\sc ii}]}

\newcommand{\oiabs}{O\,{\sc i}~}
\newcommand{\mgii}{Mg\,{\sc ii}~}
\newcommand{\civ}{C\,{\sc iv}~}
\newcommand{\cii}{C\,{\sc ii}~}
\newcommand{\heii}{He\,{\sc ii}~}

\begin{document}

\title{AEON-z5: A Candidate AGN-driven Outflow Enriching the Circumgalactic Medium at $z\simeq5.23$}

\author[0009-0007-9647-9907]{Xiaoyang Wei}
\affiliation{Department of Astronomy, Tsinghua University, Beijing 100084, China}
\email{xy-wei25@mail.tsinghua.edu.cn}

\author[0000-0001-8467-6478]{Zheng Cai}
\affiliation{Department of Astronomy, Tsinghua University, Beijing 100084, China}
\email[show]{zcai@mail.tsinghua.edu.cn}

\author[0000-0002-0427-9577]{Shiwu Zhang}
\affiliation{Department of Automation, Tsinghua University, Beijing 100084, China}
\email{zsw18@tsinghua.org.cn}

\author[0000-0002-3489-6381]{Fujiang Yu}
\affiliation{Department of Astronomy, Tsinghua University, Beijing 100084, China}
\email{yufj@mail.tsinghua.edu.cn}

\author[0000-0003-0111-8249]{Yunjing Wu}
\affiliation{Kavli Institute for the Physics and Mathematics of the Universe (WPI), The University of Tokyo Institutes for Advanced Study, The University of Tokyo, Kashiwa, Chiba 277-8583, Japan}
\email{yunjing.wu@ipmu.jp}

\author[0009-0009-4837-2165]{Shuaiyi Li}
\affiliation{Department of Astronomy, Tsinghua University, Beijing 100084, China}
\email{lishuaiy25@mails.tsinghua.edu.cn}  

\author[0000-0001-6052-4234]{Xiaojing Lin} 
\affiliation{Department of Astronomy, Tsinghua University, Beijing 100084, China}
\email{linxj21@mails.tsinghua.edu.cn}

\author[0000-0001-6251-649X]{Mingyu Li}
\affiliation{Department of Astronomy, Tsinghua University, Beijing 100084, China}
\email{lmy22@mails.tsinghua.edu.cn}

\author[0009-0005-9427-8525]{Xuelun Mei}
\affiliation{Department of Astronomy, Tsinghua University, Beijing 100084, China}
\email{mxl25@mails.tsinghua.edu.cn}

\correspondingauthor{\href{mailto:zcai@tsinghua.edu.cn}{Zheng Cai}}

\begin{abstract}
The dispersal of chemically enriched gas from galaxies into their surroundings is a key process in galaxy evolution, yet direct observational evidence at $z>5$ remains scarce. We present AEON-z5, a galaxy at $z\simeq5.23$ in the COSMOS field comprising a compact continuum-emitting core surrounded by an extended, line-dominated ionized nebula. JWST/NIRCam imaging shows that \ha+\nii~and \hb+\oiii~emission extends to projected radii of $\gtrsim4$ kpc — 2–2.5 times the typical effective radius at the host stellar mass — reaching the outer ISM and the inner CGM.
F444W grism data reveal a highly asymmetric \ha+\nii~profile, which we interpret as a bipolar outflow and decompose into three kinematic components. The dominant component has a flux-weighted velocity offset of $+489^{+48}_{-44}\ {\rm km\,s^{-1}}$ and an $\mathrm{FWHM}=630^{+86}_{-77}\ {\rm km\,s^{-1}}$, with a high-velocity wing reaching $v_{84}\sim792\ {\rm km\,s^{-1}}$. A tentative detached feature at $\Delta v_{\rm LOS}\sim2800~\mathrm{km\,s^{-1}}$, detected at $1.5\sigma$ in the 1D spectrum ($2.4\sigma$ in the 2D fit), may trace an \ha~outflow clump. These extreme kinematics are most coherently explained by the presence of an AGN in the core.
For the extended nebula, we derive a spatially resolved velocity curve whose centroid shifts progressively redward with radius, reaching $\sim200$--$400~{\rm
km~s^{-1}}$ at $r_p \sim 1.3$--$2.5$ kpc—a trend that may reflect an accelerating outflow, corotating gas, or recycled inflow viewed in projection. Crucially, the measured \nii$\lambda6584$/H$\alpha$ ratios imply near-solar N2-based abundances ($\sim 0.7-1.0\, Z_\odot$), remaining $\gtrsim0.37\,Z_\odot$ after allowing for AGN excitation and calibration systematics.
The combination of large spatial extent, substantial enrichment and extreme kinematics identifies AEON-z5 as a candidate snapshot of feedback-driven metal transport in action, offering a direct view of how early AGN activity may redistribute chemically processed gas into the CGM within the first $\sim1.1$ billion years.
\end{abstract}

\section{Introduction}
 
The dispersal of heavy elements from galaxies into their surroundings is a defining, yet observationally elusive, step of the cosmic baryon cycle. Before JWST, absorption-line studies had already established the ubiquity of \oiabs, \cii, \mgii, and \civ absorbers at $z\gtrsim5$, indicating that metals were already widespread in the intergalactic and circumgalactic medium (CGM) within the first billion years after the Big Bang \citep{2009ApJ...698.1010B,2019ApJ...882...77C,2019ApJ...883..163B,2024ApJ...963L..28Z,2024MNRAS.530.1829S}. Early imaging searches around C~\textsc{iv} absorbers at $z\sim5-6$ suggested that at least some of the responsible enrichment may have originated in relatively faint star-forming galaxies \citep{2017ApJ...849L..18C}. More recently, ALMA and \textit{JWST} observations have begun to connect metal absorption systems with individual galaxies. A [C~\textsc{ii}]-emitting galaxy was identified at a projected separation of $\simeq20$ pkpc from an \oiabs absorber at $z=5.978$, a configuration broadly consistent with stellar-feedback-driven winds transporting metals into the CGM at $z\sim5-6$ \citep{2021NatAs...5.1110W}. Similarly, the ASPIRE survey has established connections between nine metal absorbers and eight \oiii+\hb-confirmed galaxies at $z\sim6.0$--$6.5$ \citep{2024ApJ...963L..28Z}. However, absorption lines probe only one-dimensional lines of sight, providing little information on the morphology, kinematics, or ionization structure. To directly capture the ongoing transport of metals out of galaxies, one needs spatially resolved emission-line imaging and kinematics of the enriched gas beyond the host ISM.

At $z\sim2-3$, integral-field spectroscopy established the feasibility of directly imaging gaseous structures on circumgalactic scales. Narrow-band observations and MUSE and KCWI surveys revealed giant Ly$\alpha$ nebulae extending over tens to hundreds of kpc around quasars and massive galaxy overdensities \citep[e.g.][]{2014Natur.506...63C,2015Sci...348..779H,2015AAS...22531405C,2016ApJ...831...39B,2017ApJ...837...71C,2019MNRAS.482.3162A,2021MNRAS.502..494M,2024A&A...691A.210H,2026A&A...707A.380G}. Beyond Ly$\alpha$, metal emission lines such as \civ\ are crucial to revealing the co-evolution between galaxies and their large-scale gaseous environment. Ground-based integral-field surveys have already detected these lines on CGM scales around $z\sim3$ quasars, implying substantial metal enrichment within the inner halo \citep{2020ApJ...898...26G}. 
More recently, extended \civ\ and \heii\ emission with coherent inspiraling kinematics has been detected around a massive nebula at $z\sim2.3$, providing direct evidence that enriched material can return to galaxies through recycled accretion \citep{2023Sci...380..494Z}.
However, Ly$\alpha$ is subject to resonant scattering, metal lines are intrinsically faint, and at $z>3$ the rest-frame optical diagnostics lines—\ha, \nii, \oiii, and \hb—shift out of the ground-based optical window. As a result, spatially resolved constraints on the metallicity, ionization state, and dynamics of circumgalactic gas have remained largely inaccessible at $z>5$.

JWST has changed this situation. First, its infrared spectroscopy has revealed that ionized outflows and black hole (BH) accretion were already common within the first few billion years of the Universe. At $z=5.55$, GA-NIFS detected a $\gtrsim700~{\rm km\,s^{-1}}$ outflow with $\dot M_{\rm out}\sim100~M_\odot\,{\rm yr^{-1}}$ in GS\_3073 \citep{2023A&A...677A.145U}; such broad \ha~or \oiii~components with FWHM $\sim200$--$700~{\rm km\,s^{-1}}$ are seen in 30 of 130 galaxies at $z=3$--$9$ \citep{2025ApJ...984..182X}, and 40 outflow candidates among 1087 \oiii~emitters in JADES show median velocities $\simeq530~{\rm km\,s^{-1}}$ \citep{2025ApJ...994..102C}. The widespread presence of Little Red Dots and AGN further suggests that high-redshift BH accretion is far more common than classical quasar surveys implied \citep[e.g.,][]{2023ApJ...959...39H,2025ApJ...986..126K,2024ApJ...963..129M,2024ApJ...964...39G,2025A&A...697A.175S}. 
Second, extended rest-frame optical nebulae are now being resolved at $z>5$ around luminous quasars \citep{2025arXiv250907064T,2025A&A...702A.174M,2026A&A...707A.299W}, while similar detections around non-quasar hosts remain rare at these redshifts.

Methodologically, the wide-field slitless grism surveys with NIRCam---EIGER \citep{2023ApJ...950...66K}, FRESCO \citep{2023MNRAS.525.2864O}, ASPIRE \citep{2023ApJ...951L...5Y}, COSMOS-3D \citep{2025arXiv251011373M}, and SAPPHIRES \citep{2025arXiv250315587S}---provide an efficient way to identify emission-line systems without prior spectroscopic selection. NIRCam grism data have been successfully used to recover spatially resolved rest-frame optical kinematics, such as the rotation curve of a $z=5.4$ spiral galaxy \citep{2024ApJ...976L..27N}. Forward modeling further enables systematic measurements of the kinematics in rotating disk systems at $z\sim4$--$8$ \citep{2025MNRAS.543.3249D,2023arXiv231009327L}. These surveys are therefore ideally suited to search for extended ionized nebulae and to characterize their kinematics. 

In this paper, we report the discovery of AEON-z5 (An Extended Outflow Nebula at $z\simeq5.23$)\footnote{If confirmed by future observations, ``AEON'' may also be read as ``AGN-driven Extended Outflow Nebula.''} in the COSMOS field, centered at RA, Dec = $150.03664^\circ$, $2.06197^\circ$. This system directly exhibits the feedback-mediated metal transport into the outer ISM and inner CGM through extended emissions within $\sim1.1$ Gyr after the Big Bang. 
The paper is organized as follows. Section~\ref{sec:data} describes the observations and data reduction. We then examine the multi-band morphology and derive an imaging-based constraint on the \oiii/\hb~ratio in Section~\ref{sec:imaging}, followed by a detailed analysis of the F444W grism spectra in Section~\ref{sec:grism}, including forward modeling of the central kinematics, the spatially resolved velocity curve, and a tentative high-velocity clump. Section~\ref{sec:sed} presents the SED modeling of the core and nebula separately. In Section~\ref{sec:discussion}, we interpret the combined constraints on metallicity, AGN evidence, kinematic scenarios, and enrichment. Section~\ref{sec:conclusions} summarizes our main findings. 
Throughout this paper, we adopt the Planck 2018 cosmology \citep{2020A&A...641A...6P}, giving a physical scale of $1\arcsec = 6.28$ kpc at $z=5.23$. For conversions to solar units, we adopt the solar photospheric oxygen abundance of \citet{2009ARA&A..47..481A}, $12+\log(\mathrm{O}/\mathrm{H})_\odot = 8.69$. We use line wavelengths based on the CHIANTI Database \citep{1997A&AS..125..149D,2021ApJ...909...38D}\footnote{https://db.chiantidatabase.org/}.

\section{Data Reduction}\label{sec:data}
\subsection{Imaging Data}
We use deep \textit{JWST}/NIRCam observations of the COSMOS field from COSMOS-Web (GO\#1727), COSMOS-3D (GO\#5893), and SAPPHIRES (GO\#6434), spanning eight filters (F115W, F150W, F200W, F277W, F335M, F356W, F410M, and F444W). The field is also covered in the MIRI/F770W filter. 
We obtained the NIRCam data from the Mikulski Archive for Space Telescopes (MAST) at the Space Telescope Science Institute\footnote{\url{https://mast.stsci.edu/portal/Mashup/Clients/Mast/Portal.html}}. 
The specific observations analyzed can be accessed via \dataset[DOI: 10.17909/dt73-gk44]{{https://doi.org/10.17909/dt73-gk44}}.
The image reduction is based on version 1.18.1 of the standard \texttt{jwst} pipeline \citep{Bushouse2025}, with reference file \texttt{jwst\_1364.pmap}. We begin the reduction with the public uncalibrated files, then performed astrometric calibration by registering the images to \textit{Gaia} \citep{2023A&A...674A..22G} and COSMOS-Web DR1 \citep{2025A&A...704A.339S}. 
The final mosaic images are drizzled with \texttt{pix\_frac}=1.0 and a pixel size of $0.03\arcsec$.
For the MIRI F770W data, we directly use the release from \citet{harish2025cosmos}.
COSMOS HST/ACS F814W imaging data \citep{2007ApJS..172..196K} are also available.

\subsection{Grism Data}
We analyze F444W grism data from COSMOS-3D and F356W grism data from SAPPHIRES. The spectral
coverage includes \ha~and \nii$\lambda\lambda6548,6584$ in F444W, and \oiii$\lambda5007$ in F356W. The data are reduced following the procedure outlined by \citet{2023ApJ...953...53S,sun2024nircam_grism}\footnote{The code and calibration data are available at \url{https://github.com/fengwusun/nircam_grism/}}. Briefly, for each grism exposure and its corresponding short-wavelength (SW) direct image, we perform World Coordinate System (WCS) assignment, flat-field correction, super-sky background subtraction, and $1/f$ noise subtraction, particularly for the spectroscopic data. We measure astrometric offsets between the SW direct images and external imaging catalogs and then
apply them to the spectral trace model. Two-dimensional spectra are stacked for each source, and one-dimensional spectra are extracted using both boxcar and optimal extraction \citep{Horne_1986}.

We further use the semi-automated algorithm of \citet{Lin_2026} to determine the grism spectroscopic redshift by identifying emission lines in continuum-subtracted grism spectra, given a photometric-redshift prior.

\subsection{Ancillary Data}
We additionally considered archival Chandra and ALMA observations.
The ALMA data were obtained as part of the Band~6 CHAMPS program
(Project 2023.1.00180.L; PI: A. Faisst), with an effective observing
frequency of $\nu_{\rm obs}\simeq241$~GHz
($\lambda_{\rm obs}\simeq1.24$~mm). We analyzed the
primary-beam-corrected continuum image delivered by the ALMA archive
following QA2 processing, without reprocessing the calibrated
visibilities.
For the X-ray constraint, we queried the Chandra Source Catalog
version~2.1 (CSC~2.1) at the target position, using the catalog source
properties and the corresponding limiting-sensitivity estimate
\citep{2024ApJS..274...22E}.

The different properties of the galaxy's components require spatially resolved measurements. Although Subaru and IRAC data are also available, we do not include them in the analysis because of their insufficient angular resolution or sensitivity for our target. HST/WFC3 G141 grism data are also available for the target, but no clear signal is detected because of the low signal-to-noise ratio (SNR).

\section{Imaging Analysis}\label{sec:imaging}
\subsection{Multiband Morphology}
\begin{figure*}[htbp]
    \centering
    \includegraphics[width=1.\linewidth]{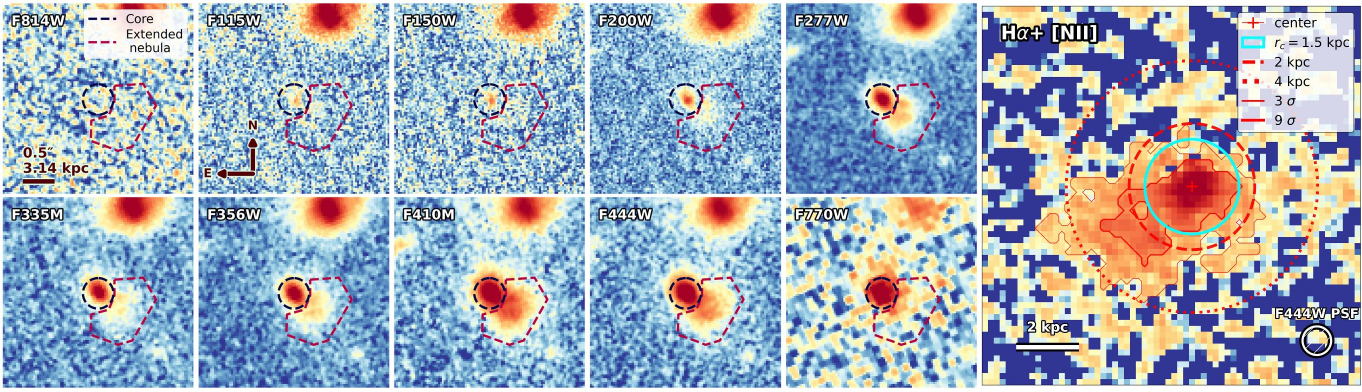}
    \caption{\textbf{Left}: Multi-band morphology of AEON-z5 in HST/ACS F814W,
JWST/NIRCam, and JWST/MIRI F770W imaging. The black dashed circle
marks the compact-core aperture ($r=1.5~{\rm kpc}$), while the red
dashed polygon marks the extended-nebula aperture. The scale bar and
compass are shown in the F814W and F115W panels, respectively. The
extended structure is most prominent in filters containing the
redshifted \hb+\oiii~and \ha+\nii~complexes, but is weak in the shorter-wavelength continuum images and
in F770W. \textbf{Right}: Spatial extent of the \ha+\nii~emission traced by the F410M$-$F444W map. This panel has been rotated
by $109^\circ$ to match the position angle (PA) of the grism exposure.
The red cross marks the adopted nebular center, defined as the peak of
the smoothed line-emission map. The thin and thick red contours trace
the $3\sigma$ and $9\sigma$ emission regions, respectively,
demonstrating that significant line emission extends beyond the
compact central component. The solid cyan circle indicates the
adopted core radius, $r_c=1.5~{\rm kpc}$, while the red dashed and
dotted circles indicate projected radii of 2 and 4 kpc, respectively.
A 2 kpc scale bar and the F444W PSF are shown at the bottom of the
right panel.
    }
    \label{fig:multiband}
\end{figure*}

The target consists of a bright compact core and an extended structure, as shown
in Figure~\ref{fig:multiband}. The extended structure is prominent in F277W, F356W,
F410M, and F444W, which contain the \hb+\oiii~and
\ha+\nii~complexes, extending to projected radii of $\gtrsim4$ kpc, but is undetected in F115W, F150W,
F200W, and F770W.

We define a compact core region with a radius of $1.5$ kpc\footnote{The radius of $1.5$ kpc corresponds to the outermost point of the $3\sigma$ contour on the side away from the extended nebula, and therefore encloses most of the flux from the compact core.} and delineate the extended region manually using \texttt{DS9}\footnote{https://sites.google.com/cfa.harvard.edu/saoimageds9} \citep{2003ASPC..295..489J}. We refer to these two regions as the core region and the nebula region, respectively, throughout this paper. The strong wavelength dependence of the extended emission, together with its counterpart seen in the grism spectrum (Section~\ref{sec:grism}), suggests that it is predominantly line-dominated. Given the similar PSF of F410M and F444W, their difference provides a map of the H$\alpha$+\nii~emission, following \citet{2024ApJ...976L..27N}. We will use this map in the spatial and spectral analyses below.

\subsection{Imaging-based \texorpdfstring{$R_3$}{R3} Constraints}\label{sec:R3}

We use the PSF-matched F277W, F335M, and F356W photometry to constrain
\hb~and the \oiii~doublet in the core and extended-nebula apertures.
The calculation models the continuum and emission-line contributions
through the filter transmission curves.
Because \hb~lies close to the F277W band edge, the inferred ratio is sensitive to the
assumed line profile and continuum slope. We therefore treat these measurements
as imaging-based excitation constraints rather than precise spectroscopic line
ratios. The full model definition, priors, and fitted fluxes are presented in
Appendix~\ref{app:R3_method}.

For the extended nebula, the whole-aperture fit gives
\begin{equation}
R_3 \equiv
\frac{F_{[\mathrm{O\,III}]\,\lambda5007}}
     {F_{\mathrm{H}\beta}}
=7.56^{+12.92}_{-3.48}.
\end{equation}
The inner and outer nebular apertures yield
$R_3=6.62^{+9.40}_{-2.94}$ and
$R_3=6.75^{+14.43}_{-3.52}$, respectively. The full results are shown in Table~\ref{tab:oiii_hbeta_profile_fits}. The two posterior distributions are broad, highly asymmetric, and strongly overlapping. We therefore find no statistically significant radial variation in \(R_3\) at the current signal-to-noise ratio. 
We infer rest-frame equivalent widths of
$\mathrm{EW}_0(\mathrm{H}\beta)=232^{+158}_{-139}\,\AA$ and
$\mathrm{EW}_0([\mathrm{O\,III}]\lambda5007)=1802^{+200}_{-205}\,\AA$
for the pure-nebula aperture. 
The large \oiii~equivalent width, although model dependent, is consistent with a scenario in which the extended nebula emission is strongly line dominated, with only a weak underlying continuum.

The core shows a moderate excitation level, with
$R_3=3.42^{+4.07}_{-1.29}$, lower than that measured in the extended
nebula. Together with the elevated \nii/\ha~ratio and the
F444W kinematics (Section~\ref{sec:grism}), this value is more suggestive of a composite/AGN-like
excitation condition than with a purely high-excitation star-forming region (see also Section~\ref{sec:AGN}). We emphasize that these imaging-based measurements are model dependent, particularly in the decomposition of \hb~near the band edge, so they serve primarily as a consistency check on the grism analysis rather than an independent spectroscopic diagnostic.

\section{Grism Analysis}\label{sec:grism}
\subsection{Emission-line Detections}
The F444W grism data reveal spatially extended \ha~and \nii~emission (Figure~\ref{fig:forward}), while the F356W data show a clear detection of \oiii$\lambda5007$ (Figure~\ref{fig:o3_spec})\footnote{The \oiii$\lambda5007$ line lies near the edge of the F356W bandpass, where the transmission is low and the uncertainties are large. We therefore use this detection only as a qualitative confirmation and do not rely on it for quantitative constraints.}.

In the F444W spectrum, the cross-dispersion direction reveals two kinematically distinct regions. The upper part, dominated by the core, exhibits a broad, multi-peaked \ha+\nii~profile with significant line broadening (Figure~\ref{fig:forward}), indicative of complex kinematics. The lower part, corresponding to the extended nebula, shows a spatially resolved velocity gradient, where the \ha~peak shifts progressively redward with increasing radius (Figure~\ref{fig:v_curve}). This contrast in kinematic behavior motivates us to treat the upper and lower regions separately. We model the upper region in two dimensions (Section~\ref{sec:forward}), while for the lower extraction, where the forward-model assumption of a spatially invariant profile breaks down, we measure the velocity centroid as a function of position by comparing the \ha~ peak in each grism row with the line-emission proxy in the direct image (Section~\ref{sec:velocity_curve}).

\subsection{Forward Modeling of the Line Emission}\label{sec:forward}
\begin{figure*}[htbp]
    \centering
    \includegraphics[width=1\linewidth]{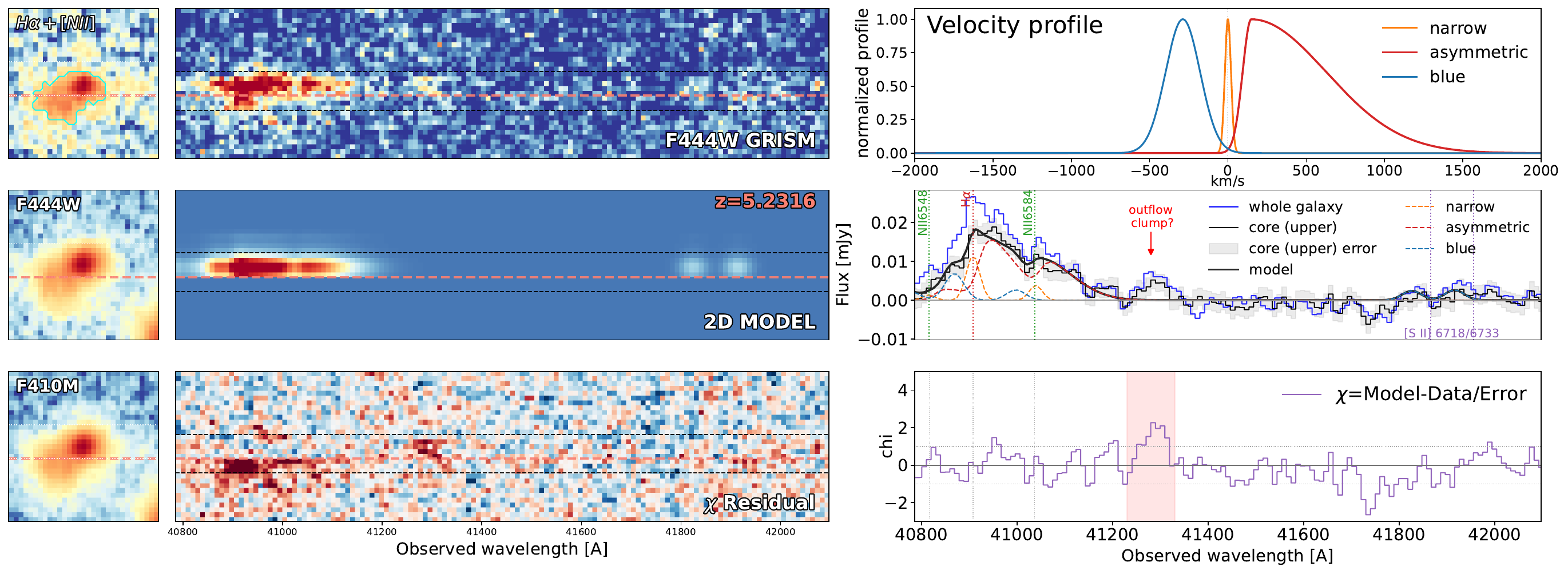}
    \caption{Forward modeling of the upper nebular component in the F444W grism data. \textbf{Left panels} show the direct-image morphology used to guide the spatial model: the continuum-subtracted F410M-F444W image tracing \ha+\nii, the F444W image, and the F410M image. The cyan contour marks the adopted source mask, and horizontal dashed lines indicate the spatial extraction aperture and upper/lower split boundary. \textbf{Middle panels} show the observed F444W 2D spectrum, the best-fit 2D forward model, and the residual in units of $\chi$. \textbf{Right panels} show the assumed velocity profiles for the narrow core, red-asymmetric split-Gaussian, and blueshifted components; the extracted 1D spectrum compared with the forward-modeled components; and the corresponding 1D $\chi$ residual. The black spectrum is the upper-region extraction, the blue spectrum is the whole-galaxy boxcar extraction, and the gray band shows the propagated $1\sigma$ uncertainty. Vertical markers indicate \ha, \nii, and \sii~wavelengths at the fitted redshift. The shaded region near 41290~\AA\ marks a positive residual discussed in Section~\ref{sec:clump}.}
    \label{fig:forward}
\end{figure*}

We forward model the F444W grism data in two dimensions. For each
kinematic component \(k\), the model assumes that the line-of-sight velocity profile is constant across the spatial extent of the source, i.e. all pixels share an identical profile, while the spatial dependence is described by the emission line map. 

Here we use the difference of the F410M and F444W images as the \ha$+$\nii~emission-line map. This map is then rotated to the grism position angle, resampled onto the native grism spatial grid. Given the similar PSF of the two bands, we do not perform additional PSF matching. We denote the resulting normalized spatial template for component \(k\) as \(T_k(y,x_{\rm d})\), where \(x_{\rm d}\) is the direct-image coordinate parallel to the grism dispersion direction. In our
fiducial model, all kinematic components use the same spatial template,
such that \(T_k=T\).

For line \(\ell\) in component \(k\), the central wavelength is
\begin{equation}
\lambda_{\ell,k}
\simeq
\lambda_{\ell,0}(1+z_{\rm ref})+\delta\lambda_k
\end{equation}
Its intrinsic profile \(P_{k,\ell}\) is convolved with the grism
line-spread function, denoted as $\widetilde P_{k,\ell}$.
Our main assumption is that
the velocity profile of each kinematic component is spatially
invariant, so 
\begin{equation}
P_{k,\ell}(\Delta\lambda;x_{\rm d},y)
\equiv
P_{k,\ell}(\Delta\lambda),
\end{equation}
After mapping \(T_k\) into the grism frame using the trace and
dispersion solution, the full model is
\begin{equation}
M(y,\lambda)
=
\sum_k\sum_\ell F_{k,\ell}
\left[
T^{\rm disp}_{k,\ell}
\ast_{\lambda}
\widetilde P_{k,\ell}
\right]
\left(y,\lambda-\lambda_{\ell,k}\right),
\end{equation}
where \(F_{k,\ell}\) is the integrated line flux. 

Motivated by the multi-peaked structure and the highly asymmetric line profiles observed for both \ha~and \nii~in the upper-region spectrum (Figure~\ref{fig:forward}), we adopt three phenomenological kinematic components, interpreted as a bipolar outflow: a narrow core component tracing the emission from the central source, a blueshifted component tracing the approaching side of the outflow, and a broad asymmetric component tracing the receding side.\footnote{A two-component model with a Gaussian narrow core and a split-Gaussian component fails to reproduce the observed profile unless the narrow-component centroid is allowed to drift significantly. When left free, the fit converges to a blue and a red split-Gaussian pair, effectively removing the central source and yielding an unphysical decomposition.} The narrow core and blueshifted components are Gaussians.\footnote{Many previous studies, \citep[e.g.]{2025ApJ...994..102C,2025ApJ...984..182X,2026arXiv260106255R} describe the line profiles of similar systems using a narrow plus a broad Gaussian component. However, those profiles are generally symmetric.} The narrow core component defines the reference redshift and has $\delta\lambda_{\rm narrow}=0$. The blueshifted component has a centroid offset $\delta\lambda_\mathrm{blue}$ and width $\sigma_{\rm blue}$. For the broad component, we use a split-Gaussian profile with independent blue-side and red-side widths, $\sigma_{{\rm out},b}$ and $\sigma_{{\rm out},r}$ and a free centroid offset $\delta\lambda_\mathrm{out}$.

For each kinematic component, the \nii~doublet ratio is fixed to the
theoretical value 
$F_{k,[\mathrm{N\,II}]\lambda6584}
=
2.96\,
F_{k,[\mathrm{N\,II}]\lambda6548}$.
The best-fit parameters are obtained by minimizing the weighted difference between the two-dimensional model and spectrum,
with
\begin{equation}
\chi^2
=
\sum_{y,\lambda}
\frac{
\left[D(y,\lambda)-M(y,\lambda)\right]^2
}{
\sigma_D^2(y,\lambda)
}
\end{equation}
where \(D(y,\lambda)\) is the observed 2D grism spectrum and
\(\sigma_D(y,\lambda)\) is the corresponding observed uncertainty. 

\begin{table*}
\centering
\caption{
Best-fit kinematic components for the upper part of the F444W grism spectrum.
Values are posterior medians with 16th--84th percentile uncertainties.
Velocities are measured relative to the fitted narrow core component.
}
\label{tab:f444w_components}
\begin{tabular}{lcccccc}
\hline
Component
& $\delta\lambda_{\rm peak}$
& $\Delta v_{\rm cen}$
& $F_{\mathrm{H\alpha}}$
& $F_{\mathrm{[NII]}\lambda6584}$
& $\sigma$
& FWHM \\
& $(\mathrm{\AA})$
& $(\mathrm{km\,s^{-1}})$
& \multicolumn{2}{c}{$(10^{-18}\,\mathrm{erg\,s^{-1}\,cm^{-2}})$}
& $(\mathrm{km\,s^{-1}})$
& $(\mathrm{km\,s^{-1}})$ \\
\hline
Narrow core
& $0$
& $0$
& $7.31^{+1.76}_{-2.50}$
& $2.54^{+1.91}_{-1.95}$
& $37^{+28}_{-20}$
& $88^{+67}_{-47}$ \\

Blue
& $-39.3^{+4.0}_{-4.0}$
& $-288^{+29}_{-29}$
& $6.46^{+2.84}_{-1.41}$
& $2.67^{+1.57}_{-1.20}$
& $109^{+66}_{-49}$
& $256^{+155}_{-115}$ \\

Asymmetric outflow
& $+22.0^{+3.4}_{-3.4}$
& $+489^{+48}_{-44}$
& $26.4^{+4.0}_{-3.2}$
& $13.9^{+2.3}_{-2.2}$
& $61^{+13}_{-12}/474^{+68}_{-59}$
& $630^{+86}_{-77}$ \\
\hline
\end{tabular}
\begin{flushleft}
\footnotesize
The systemic redshift posterior is
$z_{\mathrm{ref}}=5.2316^{+0.0006}_{-0.0004}$,
with a MAP value of $z_{\mathrm{ref}}=5.2316$.
For the narrow and blue components, a single $\sigma$ is reported because the
profiles are symmetric Gaussians. For the asymmetric outflow, the two $\sigma$
values denote the blue-side/red-side widths of the split Gaussian, and the
listed FWHM is computed as
$\mathrm{FWHM}=\sqrt{2\ln2}\,(\sigma_\mathrm{blue}+\sigma_\mathrm{red})$.
The \nii~doublet ratio is fixed with
$F_{\mathrm{[NII]}\lambda6548}
=F_{\mathrm{[NII]}\lambda6584}/2.96$.
The flux-weighted centroid of a kinematic component is
$\Delta v_{\rm cen}
=
\Delta v_{\rm peak}
+
\sqrt{\frac{2}{\pi}}
(\sigma_{v,r}-\sigma_{v,b})$,
where $\Delta v_{\rm peak}
=
c\,\delta\lambda_{\rm peak}/\lambda_{\rm sys}$
is the velocity of the profile peak relative to systemic H$\alpha$,
$\lambda_{\rm sys}
=
\lambda_{\mathrm{H\alpha},0}(1+z_{\rm ref})$,
and $\sigma_{v,b}$ and $\sigma_{v,r}$ are the blue- and red-side velocity
dispersions of the split Gaussian. For a symmetric Gaussian,
$\sigma_{v,r}=\sigma_{v,b}$, so the centroid and peak coincide.
All derived quantities and their credible intervals are evaluated
sample by sample from the posterior.
\end{flushleft}
\end{table*}

The forward modelling result is shown in Table~\ref{tab:f444w_components}. The narrow core component defines the systemic
redshift $z_{\mathrm{sys}}=5.2316$. It is relatively narrow,
with a posterior-median $\mathrm{FWHM}=88^{+67}_{-47}~\mathrm{km\,s^{-1}}$.
The blue component is offset by
$\Delta v_{\rm cen}=-288^{+29}_{-29}~\mathrm{km\,s^{-1}}$ and has
$\mathrm{FWHM}=256^{+155}_{-115}~\mathrm{km\,s^{-1}}$. The asymmetric component
contains most of the H$\alpha$ flux and has a flux-weighted centroid of
$\Delta v_{\rm cen}=+489^{+48}_{-44}~\mathrm{km\,s^{-1}}$. Its split-Gaussian
widths are highly asymmetric, with
$\sigma_{\rm blue}=61^{+13}_{-12}~\mathrm{km\,s^{-1}}$ and
$\sigma_{\rm red}=474^{+68}_{-59}~\mathrm{km\,s^{-1}}$, corresponding to
$\mathrm{FWHM}=630^{+86}_{-77}~\mathrm{km\,s^{-1}}$. For the MAP split-Gaussian profile,
the 84th-percentile velocity is approximately
$+792~\mathrm{km\,s^{-1}}$. All velocities are measured relative to the fitted
narrow H$\alpha$ component.

To estimate equivalent widths, we use the F410M photometry of the core region, which corresponds to the upper part of the spectrum. With an observed flux density of $f_{\nu,\mathrm{F410M}}=1.557\pm0.078~\mu\mathrm{Jy}$, we subtract the predicted line contribution from the spectral model, 
leaving a continuum of $0.670^{+0.089}_{-0.087}~\mu\mathrm{Jy}$ ($m_{\rm AB,cont}=24.34^{+0.15}_{-0.14}$). 
The resulting rest-frame H$\alpha$ equivalent widths are $98^{+28}_{-33}$, $87^{+42}_{-24}$, and $355^{+86}_{-66}~\AA$ for the narrow, blueshifted, and red-asymmetric components, respectively, with a summed value of $541^{+111}_{-83}~\AA$. 

The \nii$\lambda6584$/\ha\ ratios are approximately $0.35$, $0.41$, and
$0.53$ for the narrow, blueshifted, and red-asymmetric components, respectively,
while the flux-summed upper-region ratio is $\simeq0.49$. Adopting the
kinematic line profiles inferred from the forward modeling of \ha+\nii~and fitting only the \sii~fluxes, we obtain one-sided $1\sigma$ and
$3\sigma$ upper limits of $0.21$ and $0.29$, respectively, for the
flux-summed \sii$\lambda\lambda6716,6731$/\ha~ratio across all three
components. The \sii/\ha~upper limit falls below the typical threshold of $\sim0.4$ for radiative-shock-dominated emission \citep{2008ApJS..178...20A,2024A&A...690A.161L}, arguing
against shocks dominating the integrated line flux.
Together with the high \nii/\ha~ratio,
this may instead indicate high gas-phase metallicity and/or enhanced N/O,
subject to degeneracies with the ionization conditions. We defer the joint
interpretation using both \nii/\ha~and $R_3$ to
Section~\ref{sec:metal}, and the excitation mechanism to
Section~\ref{sec:AGN}.

\subsection{A Tentative High-velocity Clump}\label{sec:clump}

Figure~\ref{fig:forward} shows a detached emission feature at
$\lambda_{\rm obs}\simeq41290$~\AA, visible in the one-dimensional residual
spectrum at a significance of $\sim1.5\sigma$. A two-dimensional Gaussian
fit to the residual image gives an integrated significance of
$\sim2.4\sigma$. If interpreted as \ha~emission from the target galaxy, this
corresponds to a velocity offset of $\Delta v_{\rm LOS}\approx2800~{\rm
km\,s^{-1}}$ relative to the systemic redshift $z_{\mathrm{sys}}=5.2316$.

We investigated whether the feature could instead originate from neighboring
objects projected onto the dispersion direction.
Candidate contaminants were identified using the COSMOS-Web DR1 catalog \citep{2025A&A...704A.339S}, adopting their photometric redshifts and \texttt{mag\_auto\_f444w} magnitudes. We found five candidates (IDs 81736, 81280, 81036, 79860, and 79336) with photometric redshifts spanning $z_{\rm phot}\sim0.85$--$4.27$. Within this redshift range, the observed feature could in principle correspond to Pa$\alpha$, Pa$\beta$, [Fe\,II]~$\lambda1.257~\mu$m, He\,I~$\lambda1.083~\mu$m, or [S\,III]~$\lambda\lambda9069,9531$. These candidates have $m_{\rm F444W}=24.42$--$25.99$.
Reproducing the
observed flux for the feature would require 
rest-frame EWs of $\sim150$--$1200$~\AA, depending on the assumed transition and redshift.
Such extreme line strengths would generally require unusually intense
star formation, an AGN, or strong shocks, and would likely be accompanied by
additional detectable emission lines at predictable wavelengths.
Contamination from the currently identified neighboring sources therefore
appears unlikely.

If the feature is indeed physically associated with AEON-z5, its
continuum-subtracted equivalent width is modest,
${\rm EW}_{\rm rest}\approx21$~\AA, estimated from the F410M continuum. The inferred projected velocity,
however, is extreme, reaching
$\Delta v_{\rm LOS}\approx2800~{\rm km\,s^{-1}}$.
Such velocities are difficult to explain through gravitational motions within
the host galaxy or by normal stellar-feedback-driven winds.
Instead, the feature would be more naturally interpreted as a compact,
highly accelerated ionized-gas clump associated with an AGN-driven outflow,
or possibly a localized jet--ISM interaction.

Although this component would contain only a small fraction of the total
ionized gas mass, its very high velocity implies that it could contribute
disproportionately to the kinetic-energy budget
($\dot{E}_{\rm kin}\propto v^3$).
If confirmed by future observations, it would provide strong evidence for a
multi-phase AGN feedback process.

\subsection{Spatially Resolved Kinematics}\label{sec:velocity_curve}

\begin{figure*}[htbp]
    \centering
    \includegraphics[width=0.96\linewidth]{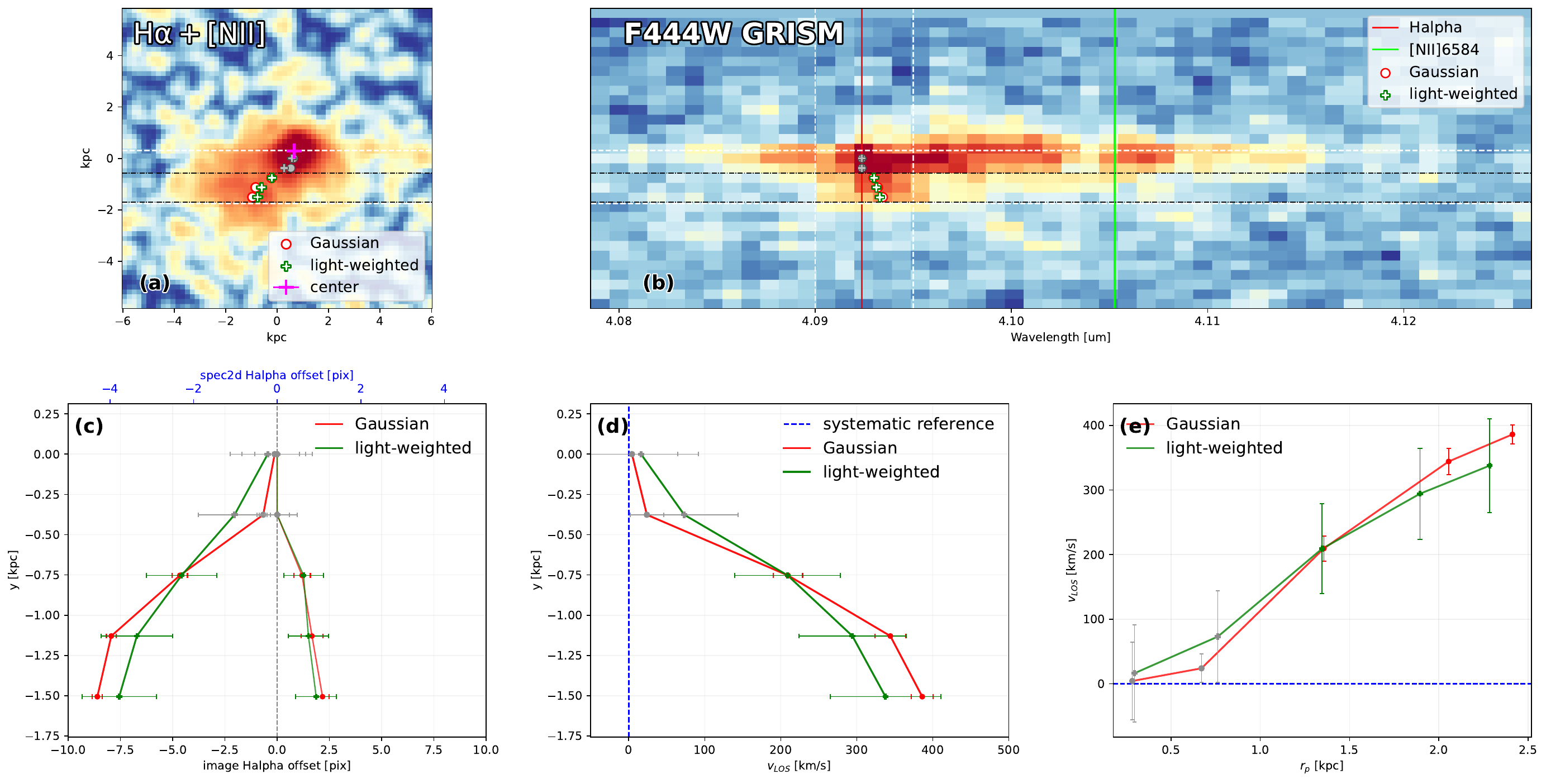}
    \caption{
Velocity curve of the extended nebula.
(a) Continuum-subtracted \ha+\nii~emission-line map constructed from the
F410M$-$F444W image. The purple cross marks the adopted nebular center, defined
as the peak of the smoothed emission-line map. Open symbols show the row-by-row
centroids measured from the direct image using a Gaussian fit or a light-weighted
centroid.
(b) F444W grism spectrum over the \ha+\nii~wavelength range. The red and green
vertical lines mark the expected wavelengths of \ha~and \nii$\lambda6584$ at
the adopted systemic redshift. Symbols show the corresponding row-by-row spectral
centroids.
(c) Direct-image centroid offsets and grism spectral centroid offsets as a
function of cross-dispersion position. The lower axis gives the centroid offset
in the emission-line map, while the upper axis gives the spectral centroid offset
in the grism frame.
(d) Inferred line-of-sight velocity curve after combining the spatial centroid
offset in the direct image with the spectral centroid offset in the grism
spectrum. The blue dashed line marks the systemic reference, defined by zero
velocity at the galaxy center.
(e) Same velocity measurements plotted against projected distance from the
nebular center. The Gaussian and light-weighted centroid methods show a consistent
increase in velocity with projected radius, reaching $\sim200$--$400~{\rm
km~s^{-1}}$ at $r_p \sim 1.3$--$2.5$ kpc.
}\label{fig:v_curve}
\end{figure*}

The upper-region forward model in Section~\ref{sec:forward} assumes a spatially invariant line profile and is therefore not applied to the visibly tilted outer emission, as shown in Figure~\ref{fig:v_curve} (b). Here we compare the position of the emission-line map with the H$\alpha$ position in each cross-dispersion row of the grism spectrum. We construct the direct-image profiles from the two image pixels corresponding to each grism row, considering the pixel scales,\footnote{The direct image has a pixel scale of $0.03''$ while the grism data have $0.06''$ per pixel. These two adjacent direct-image pixels together cover the same on-sky extent as a single grism pixel (i.e., one cross-dispersion row). We do not resample the emission-line map to the grism pixel grid, because that would degrade the resolution along the dispersion direction.} and we restrict the spectral Gaussian fits to the range $4.090$--$4.095~\mu\mathrm{m}$ to isolate the H$\alpha$ peak from \nii$\lambda6584$.

After converting the measured spatial offset in the direct image to grism-pixel units, the net wavelength displacement at a given row $y$ is given by
\begin{equation}
\Delta\lambda(y)=\Delta\lambda_{\rm grism}(y)
-\Delta x_{\rm direct}(y)\left(\frac{d\lambda}{dx}\right),
\end{equation}
where $\Delta\lambda_{\rm grism}(y)$ is the spectral centroid offset measured directly in the grism frame, $\Delta x_{\rm direct}(y)$ is the corresponding spatial centroid offset in the direct image (converted to grism pixels), and $d\lambda/dx$ is the pixel dispersion. The line-of-sight velocity is then $v_{\rm los}=c\,\Delta\lambda/\lambda_{\rm H\alpha}$, yielding the velocity curves shown in Figures~\ref{fig:v_curve} (d) and (e). The zero point is defined by the H$\alpha$ wavelength at the fitted narrow-component redshift and by the peak of the smoothed emission-line map (marked by the purple cross in Figure~\ref{fig:v_curve} (a)). Uncertainties are obtained by propagating the centroid errors measured in the direct image and grism spectrum. For Gaussian fits, the centroid error comes from the fitting covariance. For light-weighted centroids, it is estimated from the width of the emission and the number of pixels contributing to the measurement. To assess sensitivity to the profile shape, we compute the centroid along each row using two independent estimators: a Gaussian profile fit, and a light-weighted first-moment center, $x_c = \sum I_i x_i / \sum I_i$, where $I_i$ denotes the flux at pixel position $x_i$ along the row (open symbols in Figures~\ref{fig:v_curve} (a) and (b)).

Both estimators recover a larger centroid displacement at larger projected distance, reaching $\sim200$--$400~{\rm
km~s^{-1}}$ at $r_p \sim 1.3$--$2.5$ kpc (Figure~\ref{fig:v_curve} (e)). This result is nevertheless based primarily on three usable outer rows on one side of the center ($-0.12\arcsec\gtrsim y\gtrsim -0.24\arcsec$). The two rows nearest the center are excluded from the main trend because their emission is blended with the significantly broadened kinematic component of the core region, which precludes reliable centroid measurements,\footnote{So they are marked in gray color in Figure~\ref{fig:v_curve}} and the opposite side does not provide an equivalent measurement. 
As a validation of the robustness of this measurement, we repeated the analysis after subtracting the modeled core contribution from both the emission-line map and the grism spectrum in Appendix~\ref{app:vel_validation}, giving a broadly consistent result.
We will make a further discussion of the possible interpretations in Section~\ref{sec:scenarios}.

\section{SED Modeling}
\label{sec:sed}

Given the distinct physical nature of the core and the extended nebula, we model their SEDs separately using the \texttt{BEAGLE} code \citep{2016MNRAS.462.1415C,2024MNRAS.527.7217V}. We fix the redshift to $z=5.2316$, as determined from the grism forward modeling in Section~\ref{sec:forward}.

\begin{table*}
\centering
\scriptsize
\caption{
BEAGLE fitting results for the core and extended-nebula regions.
Parameter values are posterior medians with 16--84 per cent credible
intervals, and $\chi^2_{\rm min}$ is evaluated for the minimum-$\chi^2$
posterior sample. The same priors were adopted for the core and nebula
fits. Quantities marked ``derived'' were not assigned independent priors,
but were calculated from the sampled model parameters. The AGN parameters
are included only in the AGN-inclusive fits.
}
\label{tab:beagle_region1_region4}
\setlength{\tabcolsep}{3.5pt}
\begin{tabular}{lccccc}
\hline
Parameter
& Prior
& Core, with AGN
& Core, without AGN
& Nebula, with AGN
& Nebula, without AGN \\
\hline

$\log_{10}(Z_\star/Z_\odot)$
& $\mathcal{U}(-0.5,\,0.24)$
& $-0.22^{+0.25}_{-0.18}$
& $-0.24^{+0.27}_{-0.19}$
& $-0.11^{+0.22}_{-0.24}$
& $-0.12^{+0.24}_{-0.26}$ \\

$Z_\star/Z_\odot$
& induced
& $0.61^{+0.48}_{-0.21}$
& $0.58^{+0.51}_{-0.20}$
& $0.78^{+0.51}_{-0.33}$
& $0.76^{+0.56}_{-0.34}$ \\

$\log_{10}(\tau/{\rm yr})$
& $\mathcal{U}(6.0,\,10.5)$
& $6.68^{+0.49}_{-0.41}$
& $8.63^{+1.24}_{-1.39}$
& $7.12^{+2.19}_{-0.96}$
& $7.10^{+2.29}_{-0.92}$ \\

$\log_{10}(M_\star/M_\odot)$
& $\mathcal{U}(5.0,\,12.0)$
& $10.32^{+0.05}_{-0.06}$
& $9.34^{+0.06}_{-0.06}$
& $9.04^{+0.06}_{-0.05}$
& $9.04^{+0.06}_{-0.07}$ \\

$\log_{10}(Z_{\rm neb}/Z_\odot)$
& $\mathcal{U}(-0.5,\,0.24)$
& $-0.15^{+0.24}_{-0.22}$
& $-0.43^{+0.09}_{-0.05}$
& $0.07^{+0.06}_{-0.08}$
& $0.06^{+0.06}_{-0.10}$ \\

$Z_{\rm neb}/Z_\odot$
& induced
& $0.71^{+0.51}_{-0.28}$
& $0.37^{+0.08}_{-0.04}$
& $1.17^{+0.18}_{-0.19}$
& $1.15^{+0.18}_{-0.23}$ \\

$\log U_{\rm neb}$
& $\mathcal{U}(-3.95,\,-1.0)$
& $-2.52^{+0.90}_{-0.90}$
& $-2.19^{+0.21}_{-0.19}$
& $-2.04^{+0.05}_{-0.06}$
& $-2.05^{+0.06}_{-0.08}$ \\

$\log_{10}(L_{\rm acc}/{\rm erg\,s^{-1}})$
& $\mathcal{U}(43.0,\,48.0)$
& $45.61^{+0.16}_{-0.15}$
& --
& $43.79^{+0.56}_{-0.48}$
& -- \\

$\log_{10}(Z_{\rm AGN}/Z_\odot)$
& $\mathcal{U}(-0.5,\,0.24)$
& $-0.11^{+0.23}_{-0.26}$
& --
& $-0.12^{+0.24}_{-0.23}$
& -- \\

$Z_{\rm AGN}/Z_\odot$
& induced
& $0.77^{+0.54}_{-0.35}$
& --
& $0.75^{+0.55}_{-0.31}$
& -- \\

$\log U_{\rm AGN}$
& $\mathcal{U}(-4.0,\,-1.0)$
& $-3.52^{+0.22}_{-0.21}$
& --
& $-2.66^{+0.84}_{-0.81}$
& -- \\

$A_V$ [mag]
& $\tau_{V,\rm eff}\sim\mathcal{U}(0.8,\,6.0)$\tablenotemark{a}
& $0.94^{+0.28}_{-0.30}$
& $1.72^{+0.11}_{-0.10}$
& $1.30^{+0.05}_{-0.04}$
& $1.30^{+0.04}_{-0.04}$ \\

SFR$_{10}$ [$M_\odot\,{\rm yr}^{-1}$]
& derived from delayed SFH
& $<0.016$
& $190^{+31}_{-26}$
& $112^{+15}_{-13}$
& $110^{+16}_{-16}$ \\

SFR$_{100}$ [$M_\odot\,{\rm yr}^{-1}$]
& derived from delayed SFH
& $2.4\times10^{-4}\,^{+20.7}_{-2.4\times10^{-4}}$
& $23.4^{+3.5}_{-3.1}$
& $11.2^{+1.5}_{-1.3}$
& $11.0^{+1.6}_{-1.6}$ \\


$\chi^2_{\rm min}$
& --
& $3.16$
& $4.37$
& $10.07$
& $9.87$ \\
\hline
\end{tabular}

\tablenotetext{a}{
The attenuation model is CF00 with $\mu=0.3$ fixed. The sampled parameter
is the effective $V$-band optical depth,
$\tau_{V,\rm eff}\sim\mathcal{U}(0.8,6.0)$; therefore, the prior on the
reported $A_V$ is induced rather than independently uniform.
}
\end{table*}

\subsection{Model Setup and Priors}

Our fitting comprises stellar continuum, H\,{\sc ii}-region nebular emission, dust attenuation, IGM absorption, and an optional type-II AGN narrow-line region (NLR). The stellar and nebular templates are taken from the BEAGLE grid \texttt{cb2016\_n2\_mup300\_N015\_O01\_deplO70\_C100\_June16}, and the AGN/NLR component from the \texttt{AGN\_NLR} set. The photometry is measured on the images that PSF-matched with \texttt{photutils} \citep{larry_bradley_2025_14889440} to MIRI/F770W using the empirical PSFs in \texttt{stpsf} \citep{Perrin2025}. 

We parameterize the stellar population with a delayed star-formation history \citep{2003MNRAS.344.1000B},
and model dust attenuation with the Charlot \& Fall prescription \citep{2000ApJ...539..718C}.
The AGN component accounts for obscured narrow-line emission only and excludes
torus, broad-line, and shock emission \citep{2024MNRAS.527.7217V}. Motivated by the strong nitrogen lines,
we impose $\log_{10}(Z/Z_\odot)\ge-0.5$ (i.e. $Z/Z_\odot\ge 0.32$) on the stellar, nebular, and AGN
metallicities. This prior breaks part of the metallicity--attenuation degeneracy
but makes the metallicity posteriors conditional on the grism-based expectation.
For each region, we perform fits
with and without an AGN component. 
The model comparison uses $\chi^2_\text{min}$,
with the results given in
Table~\ref{tab:beagle_region1_region4} and Figure~\ref{fig:sed}.

\subsection{Core Region}

The SED fitting for the core region returns two distinct branches that produce comparable point-estimate fits ($\chi^2_{\rm min}=3.16$ for the AGN-inclusive model versus $4.37$ for the no-AGN one), yet they lead to radically different physical pictures (Table~\ref{tab:beagle_region1_region4}). The first branch describes a recent quenching event in an AGN-dominated core: the AGN-enabled solution implies a massive stellar component ($M_\star=2.1^{+0.3}_{-0.3}\times10^{10}\,M_\odot$)
and currently hosts very little ongoing star formation ($\mathrm{SFR}_{10}<0.016\,M_\odot\,\mathrm{yr}^{-1}$). The second branch, without an AGN component, instead favors a much less massive stellar population ($M_\star \simeq 2.2\times10^{9}\,M_\odot$) and compensates the same photometric SED through an extremely intense recent starburst ($\mathrm{SFR}_{10}\simeq190\,M_\odot\,\mathrm{yr}^{-1}$).

Thus, the two branches present a classic degeneracy: adopting the AGN/NLR interpretation points to a largely quiescent core in terms of ongoing star formation, whereas the alternative forces the red/line-enhanced SED to be reproduced by intense star formation without invoking any AGN contribution.

Focusing on the AGN-enabled branch, the inferred metallicities are moderately
high, with
$Z_\star=0.61^{+0.48}_{-0.21}\,Z_\odot$ and
$Z_{\rm neb}=0.71^{+0.51}_{-0.28}\,Z_\odot$, conditional on the imposed
metallicity prior. The fitted AGN/NLR component has an integrated thermal
accretion-disc luminosity of
$\log_{10}(L_{\rm acc}/{\rm erg\,s^{-1}})
=45.61^{+0.16}_{-0.15}$ and a low ionization parameter of
$\log U_{\rm AGN}=-3.52^{+0.22}_{-0.21}$. We note that the
\texttt{BEAGLE-AGN} quantity $L_{\rm acc}$ is the accretion-disc luminosity instead of the total intrinsic AGN bolometric luminosity, which may additionally include emission from the hot X-ray corona \citep{2024MNRAS.527.7217V}, though they should have the same order of magnitude. 
The ALMA Band-6 non-detection ($S_{\rm 1.24mm}<0.60~{\rm mJy}$ at $3\sigma$) provides an independent check on this SED. The BEAGLE model predicts $S_{\rm BEAGLE}(1.24~{\rm mm})\simeq0.006~{\rm mJy}$ at the ALMA wavelength, well below the observed limit. The non-detection is therefore fully consistent with the fitted SED and does not require an additional luminous dusty component in the core.
The low-ionization solution is qualitatively
consistent with the moderate \oiii/\hb\ ratio discussed in
Section~\ref{sec:R3}. All values quoted here are drawn from the full fitting
results in Table~\ref{tab:beagle_region1_region4}.

\subsection{Extended Nebular Region}
For the extended region, the minimum-$\chi^2$ values are nearly identical ($10.07$ with AGN and $9.87$ without), and both fits produce similar physical parameters (Table~\ref{tab:beagle_region1_region4}). We therefore adopt the no-AGN fit as our baseline. The no-AGN fit recovers a stellar mass of $M_\star \sim 1.1\times10^9\,M_\odot$ and a nominal star-formation rate of $\mathrm{SFR}_{10} \simeq 110\,M_\odot\,\mathrm{yr}^{-1}$, with a prior-conditional near-solar metallicity ($Z_{\rm neb} \sim 1.1\,Z_\odot$) and a higher ionization parameter ($\log U_{\rm neb} \simeq -2.05$) than in the core.

However, the stellar mass and SFR should not be interpreted as robust measurements of an underlying stellar population. The extended-region SED is dominated by strong rest-frame optical emission lines, while the \texttt{BEAGLE} nebular component is powered by ionizing photons produced by the fitted young stellar population. The model can therefore increase the recent SFR and associated stellar normalization primarily to reproduce the line luminosities, even when the observed continuum provides little independent support for such an extreme star-forming population. The systematic overprediction of the F770W flux (see Figure~\ref{fig:sed}) further suggests that the fitted stellar continuum is too strong, indicating that the current stellar templates do not adequately represent the actual stellar population in this region. A similar tension was found by \citet{2025MNRAS.542..960D} for the extended emission-line clouds around JADES-GS-518794 at $z=5.89$ (see their Section 5.3): standard star-forming fits returned low stellar masses but extremely high SFRs, placing them $\sim1.5$--$1.7$ dex above the main sequence, and the inferred parameters were accordingly treated as illustrative of model inadequacy rather than as physical measurements. Our extended-region fit shows the same behavior that \texttt{BEAGLE} formally reproduces the line-dominated SED by assigning a large recent SFR, but the resulting continuum and weak short-wavelength emission suggest that this solution is not a secure physical description of the nebula.

\section{Discussion}\label{sec:discussion}
\subsection{Gas Metallicity and Ionization}\label{sec:metal}
\begin{figure}[htbp]
    \centering
    \includegraphics[width=0.86\linewidth]{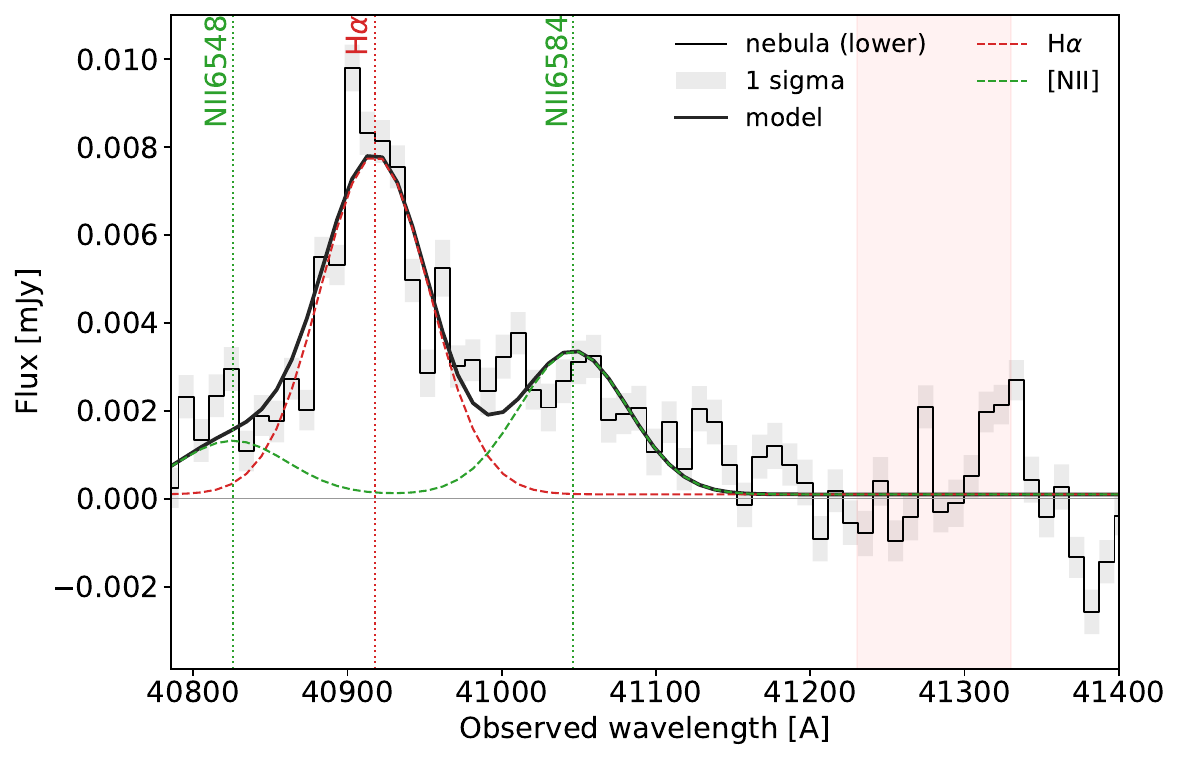}
    \caption{
    Direct 1D Gaussian fit to the lower part of the F444W grism spectrum. The black step
    curve shows the row-summed lower-aperture spectrum (marked by black dash lines in Figure~\ref{fig:v_curve}), and the gray band shows
    the propagated $1\sigma$ uncertainty. The solid dark curve is the best-fit
    model, consisting of H$\alpha$ and the \nii~doublet with a common velocity
    shift and Gaussian width. Dashed curves show the individual H$\alpha$ and
    \nii~contributions. Vertical dotted lines mark the fitted wavelengths of
    \nii$\lambda6548$, H$\alpha$, and \nii$\lambda6584$.
    }
    \label{fig:lower_Ha_NII}
\end{figure}

We next examine the ionization state and gas-phase metallicity using rest-frame optical line ratios from the F444W grism spectrum. We first consider the N2 index, $\mathrm{N2} \equiv \log(F_{\mathrm{[NII]}\lambda6584}/F_{\mathrm{H\alpha}})$. For the upper, core-dominated region, we use the three-component forward model described in Section~\ref{sec:forward}, which decomposes the emission into a narrow core component, a blueshifted Gaussian component, and a red-asymmetric component. Summing over these components yields $F_{\mathrm{[NII]}\lambda6584}/F_{\mathrm{H\alpha}} = 0.49^{+0.08}_{-0.09}$ ($\mathrm{N2} = -0.31^{+0.07}_{-0.09}$). For the lower, extended nebula region, where the SNR is lower, we fit the row-summed F444W 1D spectrum directly with a single Gaussian component shared by H$\alpha$ and the \nii~doublet, with a common wavelength shift and velocity width; the \nii$\lambda6548$ flux is tied to \nii$\lambda6584$ via the theoretical ratio of 1:2.96. Figure~\ref{fig:lower_Ha_NII} shows the fitting. From the MCMC posterior we obtain $F_{\mathrm{[NII]}\lambda6584}/F_{\mathrm{H\alpha}} = 0.326^{+0.051}_{-0.050}$ ($\mathrm{N2} = -0.486^{+0.063}_{-0.072}$).

Assuming H\,{\sc ii}-region excitation, the high-redshift AURORA N2 calibration of \citet{2026ApJ..1003..228S} yields metallicities of $Z=0.98^{+0.11}_{-0.12}\,Z_\odot$ ($0.97^{+0.59}_{-0.37}\,Z_\odot$) for the upper region, and $Z=0.76^{+0.07}_{-0.08}\,Z_\odot$ ($0.75^{+0.45}_{-0.28}\,Z_\odot$) for the lower nebula.\footnote{The values in parentheses propagate the measurement errors and the intrinsic $0.20$-dex scatter of the \citet{2026ApJ..1003..228S} calibration in quadrature.} We note that the AURORA calibration formally extends to $Z\simeq0.81\,Z_\odot$. The upper-region interval ($Z=0.86$--$1.09\,Z_\odot$) thus lies entirely beyond this range and should be treated as an extrapolation. These values quantify what the measured N2 ratios would imply under H\,{\sc ii}-region excitation and are not excitation-independent metallicity measurements. In composite systems, AGN narrow-line emission can enhance the observed N2 ratio. We adopt an illustrative AGN correction of $\Delta{\rm N2}_{\rm AGN}=0.30$ dex, which lies within the range predicted by the $z=3$ metal-rich NLR mixing models of \citet{2013ApJ...774..100K} for AGN contributions of approximately $30$--$50\%$ to the total H$\alpha$ flux. If the reported $\sim0.18$ dex enhancement in N/O at high redshift \citep{2026A&A...709A.199C} is additionally approximated as an equivalent shift in N2, the total illustrative correction becomes $\Delta{\rm N2}=0.48$ dex.
Applying this correction to the nominal N2-based abundances yields metallicities of approximately $0.48\,Z_\odot$ and $0.37\,Z_\odot$ for the upper and lower regions, respectively, indicating that the gas in both regions remains substantially enriched even under this conservative correction.

The imaging-based \oiii/\hb~ratio serves as a consistency check. For
the whole pure-nebula aperture, the fiducial result is
$R_3=7.56_{-3.48}^{+12.92}$, and the radial sub-regions yield broad, overlapping
posteriors. Under an upper-branch H\,{\sc ii}-region calibration, the median
$R_3$ would lie near or above the calibration turnover, corresponding approximately to $Z\sim0.2\,Z_\odot$, in tension with the N2 values,
while the lower end of the posterior permits values up to $\sim0.5\,Z_\odot$.
The high-$R_3$ tail lies outside the calibration range. Given the photometric
degeneracy, sensitivity to the \hb~centroid, and double-valued calibration,
we regard $R_3$ as a qualitative check rather than a precise abundance indicator.

Overall, despite these caveats, the N2 measurements consistently point to metal-enriched gas in both the core-dominated upper region and the extended lower nebula, supporting widespread enrichment across the system.

\subsection{Evidence for an Active Galactic Nucleus}
\label{sec:AGN}

Several observables, though not conclusive, jointly support the presence of an AGN in the central region.

The most compelling evidence for an AGN contribution comes from the kinematics of the non-systemic emission. The red-asymmetric component reaches an intrinsic flux-weighted velocity offset of $\simeq489~\mathrm{km\,s^{-1}}$, an FWHM of $\simeq630~\mathrm{km\,s^{-1}}$, $v_{84}\simeq792~\mathrm{km\,s^{-1}}$, and $v_{96}\sim1109~\mathrm{km\,s^{-1}}$, all measured relative to the systemic velocity. Such extreme velocities are difficult to produce by starburst-driven outflows alone. \citet{2026arXiv260106255R} find outflow velocities of $\sim170$--$600~\mathrm{km\,s^{-1}}$ in massive star-forming galaxies at $z\sim3$--$9$ with no evidence of AGN, and \citet{2025ApJ...984..182X} report $80$--$500~\mathrm{km\,s^{-1}}$ in similar systems. The $v_{84}$ and the $v_{96}$ values seen here thus lie well beyond the typical range of pure star-forming systems, pointing to an AGN-driven origin.\footnote{The $v_{84}$ here is analogous to the characteristic outflow velocities ($v_{\rm out}$ or $|v_{\rm broad}-v_\mathrm{narrow}|+\rm{FWHM}/2$) commonly reported in the literature. Because our line profile is highly asymmetric, $v_{84}$ provides a more meaningful characterization of the high-velocity wing than a symmetric FWHM.}

Beyond this kinematic signature, the SED fitting provides additional support. The AGN-inclusive model yields $\chi^{2}_{\rm min}=3.16$ versus $4.37$ for the stellar-only model under the same prior (Section~\ref{sec:sed}).

Furthermore, a tentative detached emission feature at $\Delta v_{\rm LOS}\approx2800~\mathrm{km\,s^{-1}}$ is also present (Section~\ref{sec:clump}); if confirmed as physically associated with the system, its extreme velocity would provide strong additional evidence for an AGN-driven outflow.

We assess whether the inferred AGN luminosity can plausibly power the candidate outflow. For ionized gas at $T_e\simeq10^4$ K under Case-B recombination, the H$\alpha$-derived mass is
\begin{equation}
M_{\rm ion} \simeq 3.2\times10^6 \left(\frac{L_{\rm H\alpha}}{10^{41}}\right) \left(\frac{100}{n_e}\right) M_\odot,
\end{equation}
with $n_e$ the electron density \citep{2023A&A...676A..53L}. Using the time-averaged prescription $\dot{M}_{\rm out}=C M_{\rm ion} v_{\rm out}/R_{\rm out}$ and $\dot{E}_{\rm out}= \frac{1}{2}\dot{M}_{\rm out}v_{\rm out}^{2}$ \citep{2018NatAs...2..198H}, with $C=1$, $v_{\rm out}\sim489\ {\rm km\,s^{-1}}$ (centroid velocity of the asymmetric outflow component), $R_{\rm out}\sim1.5$ kpc, and $L_{\rm H\alpha}=8.1\times10^{42}\ {\rm erg\,s^{-1}}$ (luminosity inferred from forward modelling), we obtain for $n_e=100$, $300$, and $1000\ {\rm cm^{-3}}$ ranges of $M_{\rm ion}\sim2.6\times10^8$ to $2.6\times10^7\,M_\odot$, $\dot{M}_{\rm out}\sim88$ to $8.8\,M_\odot\,{\rm yr^{-1}}$, and $\dot{E}_{\rm out}\sim6.6\times10^{42}$ to $6.6\times10^{41}\ {\rm erg\,s^{-1}}$.

The SED fit gives $L_{\rm acc}=10^{45.61}=4.1\times10^{45}\ {\rm erg\,s^{-1}}$. Adopting $L_{\rm bol}\sim L_{\rm acc}$ as an
order-of-magnitude estimate, 
the corresponding coupling efficiencies are $\epsilon_{\rm coup}=\dot{E}_{\rm out}/L_{\rm bol}\simeq1.6\times10^{-3}$,
$5.3\times10^{-4}$, and $1.6\times10^{-4}$ for the three densities. 
These small efficiencies
show that the inferred AGN is energetically capable of powering the
strongest asymmetric outflow component in our fiducial model \citep{2018NatAs...2..198H}.

If the central source is an AGN, adopting an Eddington ratio of
$\lambda_{\rm Edd}=1$ gives
$M_{\rm BH}\simeq3.3\times10^{7}\,M_\odot$
for $L_{\rm bol}\simeq4.1\times10^{45}\ {\rm erg\,s^{-1}}$,
corresponding to
$M_{\rm BH}/M_\star\simeq1.6\times10^{-3}$.
This black-hole-to-stellar mass ratio is consistent with the range observed
for early AGN hosts, although it is not unusually elevated compared with the
overmassive black-hole populations reported at $z>4$
\citep{2023ApJ...957L...3P,2024ApJ...964..154P}.

The CSC limiting-sensitivity product gives a broad-band flux limit of
$F_{0.5-7\,{\rm keV}}<1.66\times10^{-15}\,
{\rm erg\,s^{-1}\,cm^{-2}}$. Defining the X-ray bolometric correction as
$K_{\rm X}\equiv L_{\rm bol}/L_{2-10\,{\rm keV}}$, and comparing with
$L_{\rm acc}\simeq4.1\times10^{45}\ {\rm erg\,s^{-1}}$ as an order-of-magnitude
proxy for $L_{\rm bol}$, this implies $K_{\rm X}\gtrsim20$ with an assumed X-ray spectral slope of 1.6, consistent with
typical AGN bolometric corrections \citep{2020A&A...636A..73D}. The X-ray
non-detection therefore does not rule out the AGN interpretation, but only
disfavors an unusually X-ray-bright unobscured AGN.

The combined evidence most coherently points to an AGN in the core of AEON-z5.

\subsection{Possible Scenarios}\label{sec:scenarios}
\begin{figure*}[htbp]
    \centering
    \includegraphics[width=1.\linewidth]{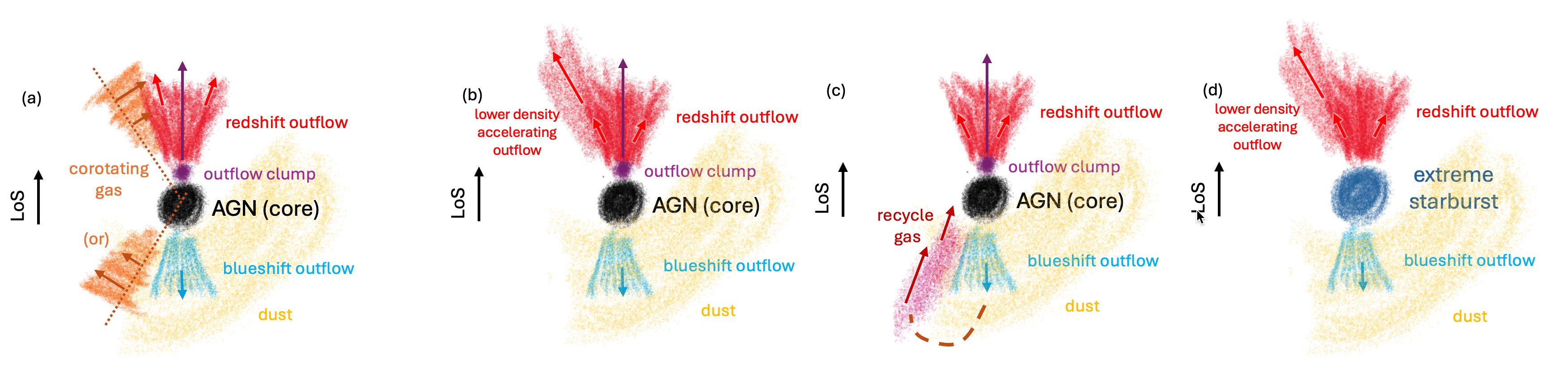}
    \caption{
Schematic illustrations of four possible interpretations of the central and
extended ionized gas. \textbf{(a)} A compact AGN-driven bipolar outflow
coexisting with gravitationally supported or corotating enriched gas. The
corotating component may appear on either the redshifted or blueshifted side
of the nucleus, depending on its position and viewing geometry.
\textbf{(b)} An AGN-driven outflow that accelerates after propagating into a
lower-density environment with reduced mass loading.
\textbf{(c)} A compact bipolar outflow accompanied by previously expelled,
bound gas that is returning toward the galaxy as part of a recycled
galactic-fountain flow.
\textbf{(d)} An extreme starburst-driven wind, with the outer gas potentially
accelerating as it expands into a lower-density medium.
}\label{fig:scenario}
\end{figure*}

The F444W spectrum reveals two kinematically distinct regions in the cross-dispersion direction: the core shows a broad, multi-peaked profile with significant line broadening, while the extended nebula exhibits a spatially resolved velocity gradient with the H$\alpha$ peak shifting redward with radius.  Although we treat these as two distinct kinematic regimes, the gas components need not be spatially separated. The overall scenario is not uniquely determined. Several interpretations remain for the rising velocity gradient at large radii, and for whether an AGN is present. We discuss these possible scenarios in turn.

\medskip
\noindent\textbf{Scenario 1: gravitationally supported outer nebula (Figure~\ref{fig:scenario}a).}
In this scenario, the central emission traces the currently 
AGN-driven bipolar outflow, while the extended redshifted nebula consists of
gas expelled during an earlier feedback episode that has subsequently slowed,
remained bound to the galaxy, and acquired corotational motion within the host potential. 
Because the observed velocity is projected along the line of sight,
corotating gas may appear on either the redshifted or blueshifted side of the
central bipolar outflow, depending on its location and viewing geometry. If the measured velocity
gradient is interpreted as circular motion, the enclosed dynamical mass can
be approximated as
\begin{equation}
M_{\rm dyn}(<r)
=
\frac{v_{\rm LOS}^{2}r}{G\sin^{2}i},
\label{eq:mdyn_rotation}
\end{equation}
where $i$ is the inclination of the rotating structure. At the outermost
measured radius, $r_p\sim2.4$ kpc, the observed velocity of
$v_{\rm LOS}\simeq340$--$385\ {\rm km\,s^{-1}}$ implies an
inclination-uncorrected lower limit of
$M_{\rm dyn}(<2.4\,{\rm kpc})\simeq(6.5$--$8.3)\times10^{10}\,M_\odot$,
or $\log(M_{\rm dyn}/M_\odot)\simeq10.81$--$10.92$. This is approximately
$3$--$4$ times the total SED-derived stellar mass,
$\log(M_\star/M_\odot)=10.32$, and would increase by a factor
$1/\sin^{2}i$ after inclination correction. Consider this correction, a similarly dynamical
mass relative to the stellar mass has been inferred for the rapidly rotating
FRESCO galaxy at $z=5.4$ \citep{2024ApJ...976L..27N}, demonstrating that massive,
rapidly rotating systems can already exist at this epoch.

The dynamical timescale of the outermost gas is $t_{\rm dyn}\sim R/v\sim6~\mathrm{Myr}$.
Proximity-zone measurements indicate a broad distribution of individual quasar on-times, ranging from below $\sim1~\mathrm{Myr}$ to beyond $\sim30~\mathrm{Myr}$ \citep{2019MNRAS.484.3897K,2021MNRAS.505.5084W}. A dynamical time of $\sim6~\mathrm{Myr}$ is therefore broadly consistent with one or more coherent luminous AGN episodes.

Nevertheless, the present source differs from a regular rotating disk.
The detected nebula is predominantly located on one side of the nucleus and
shows a largely one-sided redshifted velocity gradient rather than a clearly
antisymmetric rotation pattern. The gravitationally supported interpretation
therefore requires either that the counter-rotating side is obscured or too
faint to detect, or that the emitting gas occupies an asymmetric stream,
warped structure.

\medskip
\noindent\textbf{Scenario 2: accelerating outflow
(Figure~\ref{fig:scenario}b).}
In this scenario, the central emission traces the dense 
AGN-driven bipolar outflow, while the extended redshifted nebula consists of lower-density accelerating ionized outflow gas.

Radially increasing velocities have been reported in nearby AGN outflows.
\citet{2025NatAs...9..907M} find that some systems accelerate beyond
$\sim1$ kpc, which they interpret as a transition from a momentum-driven to
an energy-conserving flow. \citet{2025A&A...703A.230Z} suggest that acceleration can
also occur after the outflow leaves the dense center and enters a lower-density
region where the swept-up mass increases more slowly. A similar mechanism
could operate here. If the extended nebula resides in a lower ambient density or column-density environment, the same AGN wind or radiation field can accelerate it more efficiently. In particular, a declining ambient density reduces the mass loading of the expanding flow, while lower-column clouds experience a larger radiative acceleration per unit mass.

Future IFU observations would be essential to verify this scenario, by revealing a continuous position--velocity connection and enabling a biconical model that matches both the velocity rise and the one-sided morphology.

\medskip
\noindent\textbf{Scenario 3: recycled inflow in a galactic-fountain cycle
(Figure~\ref{fig:scenario}c).}
In this scenario, the compact central emission traces the currently active
AGN-driven bipolar outflow, while part of the extended nebula consists of
previously expelled, enriched gas that remained bound and is now returning
toward the galaxy. Because recycled gas may follow curved, inspiraling
trajectories with both radial and tangential velocity components, its
projected line-of-sight velocity can increase or remain roughly constant with projected distance from
the galaxy. Such a profile was observed around MAMMOTH-1 and reproduced by
a model of enriched inspiraling streams \citep{2023Sci...380..494Z}.
This interpretation places the extended nebula within a
galactic-fountain cycle, in which feedback ejects metal-enriched ISM gas into
the surrounding halo before part of it cools and reaccretes. Such recycling
can redistribute metals through the outer ISM and inner CGM while also
returning previously processed material to the galaxy, thereby linking AGN
feedback, circumgalactic enrichment, and the continued growth of the host.

\medskip
\noindent\textbf{Scenario 4: starburst-driven outflow (Figure~\ref{fig:scenario}d).}
The no-AGN SED solution allows a high nominal SFR of
${\rm SFR}_{10}\simeq190\,M_\odot\,{\rm yr^{-1}}$, so an extreme dusty starburst cannot be fully ruled out as the ionizing source and driver of the observed non-systemic motions. For continuous star formation, stellar winds and supernovae provide a mechanical power of order
$\dot E_{\rm SF}\sim7\times10^{41}({\rm SFR}/M_\odot\,{\rm yr^{-1}})
\simeq1.3\times10^{44}\ {\rm erg\,s^{-1}}$
\citep{1999ApJS..123....3L,2005ARA&A..43..769V}. For
$v_{\rm out}\sim800\ {\rm km\,s^{-1}}$, $R_{\rm out}=1.5$ kpc, and the
H$\alpha$-derived ionized mass adopted above, the warm ionized component
requires $\dot M_{\rm out}\simeq141(100\,{\rm cm^{-3}}/n_e)\,
M_\odot\,{\rm yr^{-1}}$ and
$\dot E_{\rm out}\simeq2.9\times10^{43}
(100\,{\rm cm^{-3}}/n_e)\ {\rm erg\,s^{-1}}$. The corresponding mass-loading
factor is approximately $0.74$, $0.25$, and $0.07$ for $n_e=100$, $300$, and
$1000\ {\rm cm^{-3}}$, respectively. An extreme compact starburst is therefore not energetically excluded, although it would require a relatively efficient conversion of stellar feedback into warm ionized gas.

Nevertheless, the extreme velocities discussed in Section~\ref{sec:AGN}—$v_{84}\sim792\ {\rm km\,s^{-1}}$ and $v_{96}\sim1109\ {\rm km\,s^{-1}}$ are at/beyond the high end of what is typically observed in star-formation-driven winds at high redshift, difficult to reconcile with starburst-driven winds, as the observed ranges in pure star-forming systems at similar redshifts are only $\sim80$--$600\ {\rm km\,s^{-1}}$ \citep{2025ApJ...984..182X,2026arXiv260106255R}. Furthermore, the observed increase of velocity with radius requires additional conditions—such as declining ambient density, reduced mass loading, or continued acceleration by a pressure-driven superbubble after breakout—that are less naturally expected than in an AGN-driven energy-conserving outflow.
The candidate $\sim2800\ {\rm km\,s^{-1}}$ component cannot be accounted for by starburst feedback alone; if it were physically associated with the system, it would pose a serious challenge to this scenario. However, given its tentative status, we do not use it as a primary constraint here.

Thus, while a starburst origin for the moderate-velocity component cannot be fully excluded, the combined evidence still favors the AGN-driven scenario.

\subsection{Outflows Enriching the Outer ISM/CGM}

Figure~\ref{fig:multiband} shows that nitrogen- and oxygen-bearing ionized gas extends beyond the compact core, and Section~\ref{sec:metal} indicates that metal-enriched material is present in the outer regions of the system. Here we examine the physical scale of this metal-line emission relative to the stellar component.

The right panel of Figure~\ref{fig:multiband} presents the spatial extent of the H$\alpha$+\nii~proxy derived from F410M$-$F444W. The $9\sigma$ contour reaches projected radii of $\gtrsim4$ kpc from the adopted center, demonstrating that metal-line-emitting gas extends well beyond the central aperture.

To place this scale in context, we compare with recent JWST mass--size relations at $z\sim5$ for the same stellar mass. These relations imply typical continuum effective radii of approximately $1.6$--$2.1$ kpc \citep{2024ApJ...963....9M,2024ApJ...962..176W,2025ApJ...988..196M,2025A&A...698A..30A}. The line emission therefore extends to roughly $2$--$2.5$ times a typical $R_e$ at this mass, placing it well into the outer-galaxy regime. For main-sequence galaxies at comparable redshift, \citet{2026ApJ..1000..308W} adopt an ISM/CGM demarcation of $\lesssim3$ kpc (ISM) and $\sim5$--$10$ kpc (CGM) based on ALPINE-CRISTAL-JWST IFU data; our $\gtrsim4$ kpc line emission therefore reaches the outer ISM and the lower boundary of the CGM at $z\sim5$. Adopting the lower stellar mass from the no-AGN model would predict an even smaller $R_e$, making the contrast between the line extent and the stellar half-light radius even larger. The conclusion that the line emission reaches well beyond the typical stellar scale is thus qualitatively robust to the choice of SED model.

The metallicity of the extended gas provides an additional constraint on its origin. For the AGN-model stellar mass, $\log(M_\star/M_\odot)=10.32$, recent JWST-based mass--metallicity relations at $z\sim5$ predict typical ISM metallicities of $Z\sim0.5$--$0.8\,Z_\odot$ \citep{2023ApJS..269...33N,2025ApJ...978..136S}. The N2-based metallicities inferred for the extended regions, $Z\simeq0.75\,Z_\odot$ (lower nebula), are therefore comparable to the expected enrichment level of the host ISM. The gas detected at several kpc is thus clearly
chemically processed rather than pristine or weakly enriched material.

Such enrichment at large radius is naturally expected if enriched
gas has been redistributed through the outer galaxy. When considered together with the AGN-like central kinematics and the energetically viable ionized outflow discussed above, the large spatial extent and substantial enrichment support a picture in which AGN-driven feedback transports and mixes chemically processed material into the outer ISM and potentially the inner CGM.

\section{Conclusions}
\label{sec:conclusions}

We have presented JWST/NIRCam imaging, slitless-grism spectroscopy, and SED
modeling of an extended ionized nebula surrounding
AEON-z5 at $z\simeq5.23$. Our main conclusions are as
follows.

\begin{enumerate}

\item The system contains a compact continuum-emitting core surrounded by a
line-dominated nebula detected in filters containing
\hb+[O\,{\sc iii}] and H$\alpha$+\nii. The
H$\alpha$+\nii~ proxy extends to projected radii of
$\gtrsim4$ kpc, corresponding to approximately $2$--$2.5$ times the
typical continuum effective radius expected at the AGN-model stellar mass,
$\log(M_\star/M_\odot)=10.32$. The detected emission therefore reaches the
outer-galaxy regime and approaches the transition between the outer ISM and
inner CGM.

\item For the F444W grism analysis, forward modeling of the central H$\alpha$+\nii\ emission requires three kinematic components. The narrow reference component defines the systemic redshift at $z_{\rm ref}=5.2316^{+0.0006}_{-0.0004}$, with an $\mathrm{FWHM}\simeq88~{\rm km\,s^{-1}}$. A blueshifted component is offset by $\Delta v_{\rm cen}\simeq-288~{\rm km\,s^{-1}}$ with $\mathrm{FWHM}\simeq256~{\rm km\,s^{-1}}$. The dominant component is strongly red-asymmetric, contains most of the H$\alpha$ flux, and has a flux-weighted velocity offset of $\Delta v_{\rm cen}\simeq+489~{\rm km\,s^{-1}}$, an $\mathrm{FWHM}\simeq630~{\rm km\,s^{-1}}$, and a high-velocity wing reaching $v_{84}\simeq792~{\rm km\,s^{-1}}$ and $v_{96}\sim1109~{\rm km\,s^{-1}}$. A tentative detached feature at $\Delta v_{\rm LOS}\sim2800~{\rm km\,s^{-1}}$ may trace an outflow clump, though its significance is low ($\sim1.5\sigma$ in 1D spectrum and $\sim 2.4\sigma$ in the 2D fit). The lower, extended nebula part of the spectrum shows a distinct, predominantly one-sided velocity structure, with the centroid shifting progressively redward with projected radius to $\simeq340$--$385~{\rm km\,s^{-1}}$ at $r_p\simeq2.4$ kpc.

\item The measured \nii$\lambda6584$/H$\alpha$ ratios imply that
substantially enriched gas is present in both the central and extended
regions. Under the adopted high-redshift N2 calibration, the nominal
abundances are approximately $Z\simeq0.98\,Z_\odot$ and
$0.76\,Z_\odot$ for the upper and lower regions, respectively. Although the
absolute values remain uncertain because of possible AGN excitation,
enhanced N/O, calibration scatter, and modest extrapolation, illustrative
corrections still allow $Z\gtrsim0.37\,Z_\odot$. 

\item The combined SED and kinematic evidence favors the presence of an
AGN in the compact core. The AGN-inclusive SED model gives a lower minimum
$\chi^2$ than the no-AGN solution and yields
$\log(L_{\rm acc}/{\rm erg\,s^{-1}})=45.61$. For the fiducial asymmetric
component, the inferred ionized-gas mass, mass-outflow rate, and kinetic
power span
$M_{\rm ion}\sim2.6\times10^7$--$2.6\times10^8\,M_\odot$,
$\dot M_{\rm out}\sim8.8$--$88\,M_\odot\,{\rm yr^{-1}}$, and
$\dot E_{\rm out}\sim6.6\times10^{41}$--$6.6\times10^{42}\,
{\rm erg\,s^{-1}}$ for $n_e=1000$--$100~{\rm cm^{-3}}$. Adopting
$L_{\rm bol}\sim L_{\rm acc}$ gives coupling efficiencies of
$\sim1.6\times10^{-4}$--$1.6\times10^{-3}$, showing that the inferred AGN
can readily power the strongest asymmetric outflow component.

\item The physical origin of the extended velocity structure remains uncertain. Gravitationally supported gas, an AGN-driven outflow into lower-density environment, recycled inflow, and a starburst-driven wind remain plausible to varying degrees. Nevertheless, the central kinematics and other evidence favor an AGN-associated interpretation over the alternatives.

\item Taken together, the large spatial extent, substantial metallicity,
central kinematics, and viable feedback energetics support a picture
in which AGN-driven outflows have redistributed chemically processed gas from
the central galaxy into the outer ISM and the inner CGM. AEON-z5
therefore provides a compelling direct early example of AGN feedback transporting
metals to kpc scales and contributing to the rapid enrichment of the
circumgalactic environment at $z>5$.

\end{enumerate}

Spatially resolved IFU spectroscopy, together with an independent systemic-redshift measurement from cold gas and a direct continuum-size measurement, will be required to determine the gas geometry and distinguish the possible scenarios.

\begin{acknowledgments}
This work was supported by the National Key R\&D Program of China (grant No. 2023YFA1605600), the National Natural Science Foundation of China (grant No. 12525303), and the Tsinghua University Initiative Scientific Research Program (Z.C.). J.W. was funded by the National Natural Science Foundation of China under grant Nos. 62222508 and 62525506. Z.C. and J.W. were funded by New Cornerstone Science Foundation through the XPLORER PRIZE.

This work is based on observations made with the NASA/ESA/CSA James Webb Space Telescope. The data were obtained from the Mikulski Archive for Space Telescopes at the Space Telescope Science Institute, which is operated by the Association of Universities for Research in Astronomy, Inc., under NASA contract NAS 5-03127 for JWST. These observations are associated with program \#1727, \#5893, \#6434. The authors acknowledge these program teams for developing their observing program with a zero-exclusive-access period. 
This paper makes use of the following ALMA data 
ADS/JAO.ALMA\#2023.1.00180.L.
ALMA is a partnership of ESO (representing its member states), NSF (USA)
and NINS (Japan), together with NRC (Canada), NSTC and ASIAA (Taiwan),
and KASI (Republic of Korea), in cooperation with the Republic of Chile.
The Joint ALMA Observatory is operated by ESO, AUI/NRAO and NAOJ.
The scientific results reported in this article are based in part on data
obtained from the Chandra Data Archive and on products from the Chandra
Source Catalog version~2.1.
\end{acknowledgments}

\facilities{JWST (NIRCam, MIRI), HST (ACS), ALMA, Chandra}

\appendix

\section{Photometric Forward Modeling of \texorpdfstring{$R_3$}{R3}}\label{app:R3_method}

We jointly fit the PSF-matched F277W, F335M, and F356W measurements to
separate the \hb~and \oiii~line contributions from a weak continuum.
For each filter $b$, the model band-averaged flux density is
\begin{equation}
\begin{split}
F^{\rm model}_{\nu,b}
&=C_{335}\left\langle
\left(\frac{\lambda}{\lambda_{335}}\right)^k
\right\rangle_b
+F_{\mathrm{H}\beta}K_{b,\mathrm{H}\beta}(z)\\
&\quad+F_{[\mathrm{O\,III}]\,5007}
\left[K_{b,5007}(z)+\frac{1}{2.96}K_{b,4959}(z)\right],
\end{split}
\end{equation}
where $C_{335}$ is the continuum normalization at $\lambda_{335}$, $k$ is
the continuum slope in $F_\nu$, and the \oiii~doublet ratio is fixed to
$F_{5007}/F_{4959}=2.96$. The line-to-band conversion factor is
\begin{equation}
K_{b,i}(z)=
\frac{\int P_i(\lambda,z)T_b(\lambda)\lambda\,d\lambda}
{c\int T_b(\lambda)\lambda^{-1}\,d\lambda}\frac{1}{10^{-23}},
\end{equation}
and the continuum average is
\begin{equation}
\left\langle\left(\frac{\lambda}{\lambda_{335}}\right)^k\right\rangle_b
=
\frac{\int T_b(\lambda)\lambda^{-1}
(\lambda/\lambda_{335})^k\,d\lambda}
{\int T_b(\lambda)\lambda^{-1}\,d\lambda}.
\end{equation}
Here $P_i(\lambda,z)$ is a normalized observed-frame line profile and
$T_b(\lambda)$ is the filter throughput. Dust attenuation of the continuum is
absorbed into the local power-law parameterization. The differential extinction
between \hb~and \oiii~is small (approximately 0.04 dex for the adopted
attenuation range).

We assume that the \hb+\oiii-emitting gas shares the projected
kinematic phase traced by \ha+\nii~at the same sky position. For the
core region, we adopt the composite blue+narrow+red-asymmetric profile from
Section~\ref{sec:forward}. For the extended-nebula apertures, we use a Gaussian
profile with $\sigma_v=300~\mathrm{km\,s^{-1}}$ and a centroid offset of
$+350~\mathrm{km\,s^{-1}}$ relative to $z_{\mathrm{sys}}=5.2316$, corresponding to
$z_{\rm line}=5.2388$. We also impose the Case-B constraint
$F_{\mathrm{H}\beta}<F_{\mathrm{H}\alpha}/2.86$ using the F444W grism
measurement. The four fitted parameters are
$F_{[\mathrm{O\,III}]\,5007}$, $F_{\mathrm{H}\beta}$, $C_{335}$, and $k$.
Because only three photometric bands are available, we adopt
$0.7<k<1.3$ and sample the posterior with \texttt{emcee}
\citep{2013PASP..125..306F}.

\begin{table*}
\centering
\caption{Independent-flux photometric fitting results at
$z_{\mathrm{sys}}=5.2316$. Line fluxes are in units of
$10^{-18}\,\mathrm{erg\,s^{-1}\,cm^{-2}}$, $C_{335}$ is in nJy, and
$R_3\equiv F_{[\mathrm{O\,III}]\,5007}/F_{\mathrm{H}\beta}$. Values are
posterior medians with 16th--84th percentile uncertainties.}
\label{tab:oiii_hbeta_profile_fits}
\begin{tabular}{@{}lccccc@{}}
\toprule
Region & $F_{[\mathrm{O\,III}]\,5007}$ & $F_{\mathrm{H}\beta}$ & $R_3$ & $C_{335}$ & $k$ \\
\midrule
Core
& $69.85_{-5.86}^{+6.53}$ & $20.45_{-10.34}^{+9.84}$
& $3.42_{-1.29}^{+4.07}$ & $603.65_{-4.11}^{+4.10}$
& $0.97_{-0.18}^{+0.17}$ \\
Whole nebula
& $80.17_{-7.63}^{+6.81}$ & $10.61_{-6.37}^{+7.35}$
& $7.56_{-3.48}^{+12.92}$ & $250.09_{-7.34}^{+7.53}$
& $1.03_{-0.21}^{+0.19}$ \\
Nebula, $r<2.5~\mathrm{kpc}$
& $31.90_{-3.29}^{+3.08}$ & $4.81_{-2.64}^{+2.96}$
& $6.62_{-2.94}^{+9.40}$ & $77.12_{-3.05}^{+3.18}$
& $1.02_{-0.21}^{+0.19}$ \\
Nebula, $r\geq2.5~\mathrm{kpc}$
& $46.93_{-6.48}^{+5.05}$ & $6.97_{-4.54}^{+5.69}$
& $6.75_{-3.52}^{+14.43}$ & $174.13_{-6.45}^{+6.48}$
& $1.03_{-0.20}^{+0.18}$ \\
\bottomrule
\end{tabular}
\end{table*}

The dominant systematic is the placement of \hb~near the F277W band
edge. Small changes in the assumed centroid or profile therefore alter the
inferred \hb~contribution and broaden the effective uncertainty in $R_3$.
In addition, the extended nebula has a position-dependent centroid
(Section~\ref{sec:velocity_curve}) that is approximated here by a single
profile. These limitations motivate the conservative interpretation adopted in
Section~\ref{sec:R3}.

\section{Validation of Spatially Resolved Kinematics}\label{app:vel_validation}
\begin{figure}[htbp]
    \centering
    \includegraphics[width=0.96\linewidth]{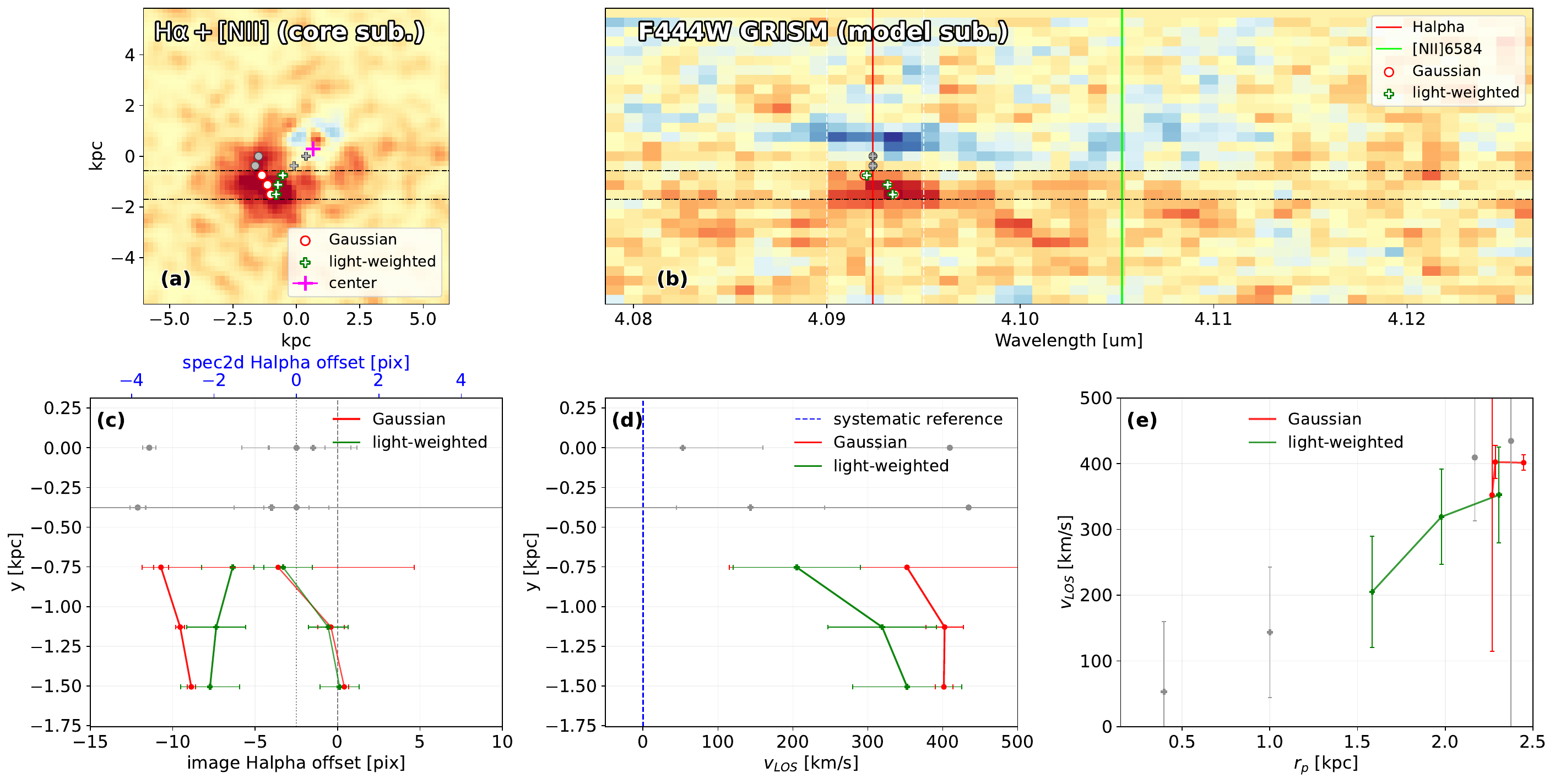}
    \caption{
Validation of the spatially resolved velocity curve after subtracting the compact Sérsic-shaped contribution. The panels and symbols follow Figure~\ref{fig:v_curve}. In panel (a), a Sérsic model fitted to the compact region is subtracted from the F410M$-$F444W emission-line map. In panel (b), the same Sérsic morphology is forward modelled through the full best-fitting three-component kinematic profile, and subtracted across the F444W grism spectrum. In panels (c)--(e), the inner two points are excluded because their centroids are particularly sensitive to the subtraction of the compact component. The light-weighted centroids recover a velocity trend consistent with the original measurement, whereas the Gaussian centroids of the three outer points show a nearly flat velocity curve, with \(v_{\rm LOS}=402\pm11~{\rm km\,s^{-1}}\) over the sampled radial range.
}\label{fig:v_curve_validation}
\end{figure}
As a validation of the velocity-curve measurement, we repeated the analysis after subtracting the compact Sérsic contribution from both the emission-line map and the grism spectrum. We fitted a Sérsic model to the core region of the F410M$-$F444W map and subtracted the full model profile, including its wings beyond the fitting aperture. For the grism spectrum, the same Sérsic morphology was forward-modeled through the full best-fitting three-component kinematic profile, and then subtracted across the F444W grism spectrum.

The results are shown in Figure~\ref{fig:v_curve_validation}. The light-weighted centroids derived from the core-subtracted data recover a velocity trend consistent with the original measurement. The Gaussian-centroid measurements of the three outer points, in contrast, show a smaller velocity gradient and are more consistent with a flat velocity curve. A constant fit gives $v_{\rm LOS}=402\pm11~{\rm km\,s^{-1}}$ across the sampled radial range. The absence of a statistically significant gradient does not exclude the accelerating outflow scenario, and even a flat profile remains compatible with the physical scenarios discussed in Section~\ref{sec:scenarios}. The validation is therefore broadly consistent with the original measurement.

\section{Supplementary Figures}
\begin{figure}[htbp]
    \centering
    \includegraphics[width=0.5\linewidth]{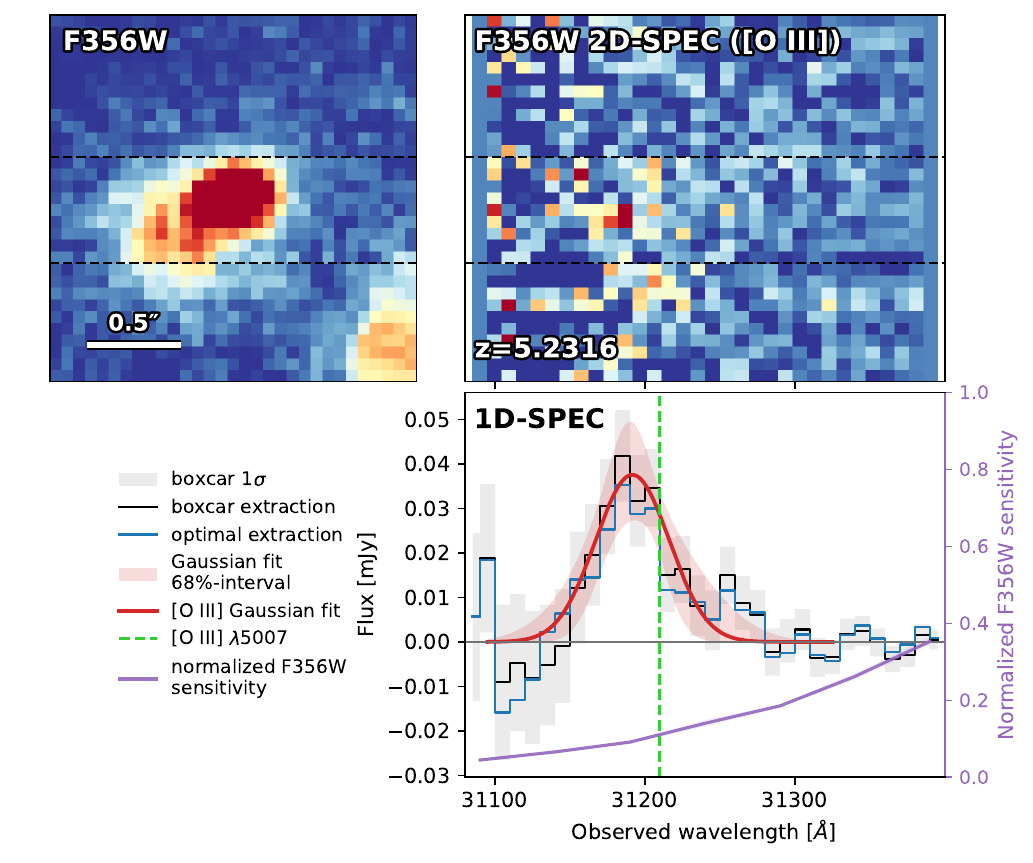}
    \caption{The upper-left panel shows the F356W direct image with a $0.5^{\prime\prime}$ scale bar.
    The upper-right panel presents the continuum-subtracted F356W two-dimensional spectrum around \mbox{[O\,III]~$\lambda5007$} at $z=5.2316$.
    The lower-right panel shows the corresponding boxcar and optimal one-dimensional extractions.
    The gray band denotes the propagated $1\sigma$ boxcar uncertainty.
    The purple curve shows the WFSS sensitivity normalized to its full-band peak, illustrating the low sensitivity near \mbox{[O\,III]}.
    The red curve and shaded region show the Gaussian fit and its 68\% Monte Carlo interval, yielding
    $F_{\mathrm{[O\,III]}} = 7.34^{+1.84}_{-1.59}\times10^{-17}\,
    \mathrm{erg\,s^{-1}\,cm^{-2}}$. Given the low transmission of the F356W filter near the \oiii~line, this flux measurement is subject to large uncertainties and is not used in the subsequent analysis.}
    \label{fig:o3_spec}
\end{figure}

\begin{figure*}[htbp]
    \centering
    \includegraphics[width=1.\linewidth]{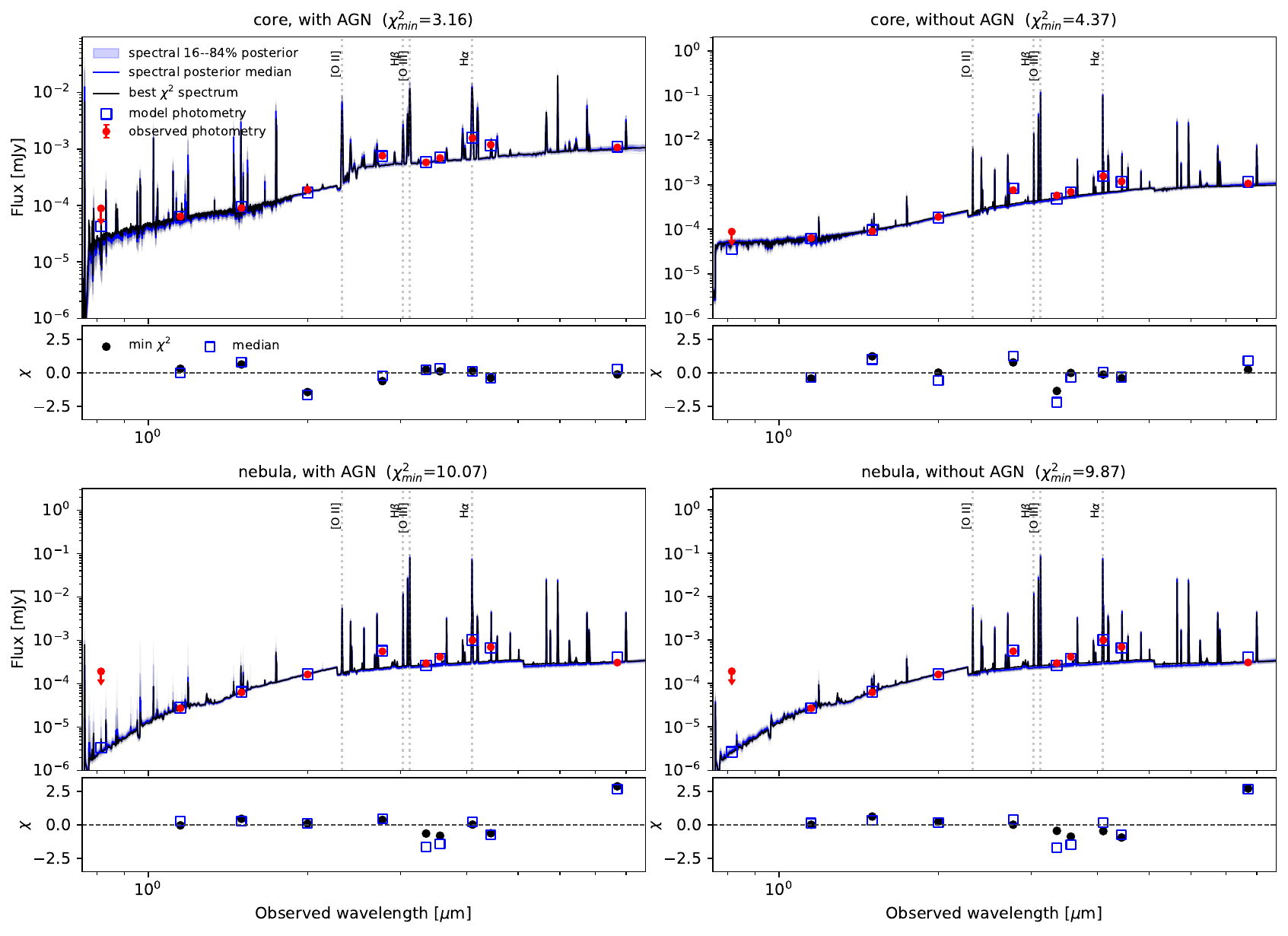}
    \caption{
    BEAGLE SED fits for the core (top) and extended nebula (bottom), each shown
    with (left) and without (right) an AGN/NLR component. In each panel, the red points show the observed photometry, while the downward red arrow at HST/F814W marks the observed $3\sigma$ upper limit. Open blue squares are the model photometry,
    the black curve is the best-$\chi^2$ spectrum, and the blue band is the
    16th--84th percentile spectral posterior; the lower strip shows the residuals
    ($\chi$) for the median (open squares) and best-$\chi^2$ (filled) models.
    Prominent rest-frame optical lines are labeled. The minimum $\chi^2$ of each
    fit is given in the panel title. For the core, the AGN model is favored
    ($\chi^2_{\rm min}=3.16$ versus $4.37$); for the
    nebula the minimum-$\chi^2$ values are similar and the evidence mildly
    favors the model without an AGN component.}
    \label{fig:sed}
\end{figure*}

Figure~\ref{fig:o3_spec} shows the F356W grism spectrum with \oiii~detection. Figure~\ref{fig:sed} presents the SED fitting results for both core and nebula regions, with and without AGN model.

\bibliography{reference}
\bibliographystyle{aasjournalv7}

\end{document}